\documentclass[twocolumn,traditabstract,longauth]{aa}

\usepackage{graphicx}
\usepackage{txfonts}
\usepackage{xcolor}
\usepackage{comment}
\usepackage[breaklinks, colorlinks, citecolor=blue]{hyperref}
\definecolor{pink}{rgb}{0.5,0.,0.5}

\newcommand{\gn}{GOODS-North}
\newcommand{\xrms}{cross-variance}

\begin{document}

\title{NIKA2 Cosmological Legacy Survey. First measurement of the confusion noise at the IRAM 30\,m telescope}

   \author{N.~Ponthieu
          \inst{\ref{IPAG},\ref{LAM}}
          \and
          F.-X.~D\'esert \inst{\ref{IPAG}}
          \and
          A.~Beelen \inst{\ref{LAM}}
         \and
         R.~Adam \inst{\ref{OCA}}
    \and  P.~Ade \inst{\ref{Cardiff}}
    \and  H.~Ajeddig \inst{\ref{CEA}}
    \and  S.~Amarantidis \inst{\ref{IRAME}}
    \and  P.~Andr\'e \inst{\ref{CEA}}
    \and  H.~Aussel \inst{\ref{CEA}}
    \and  A.~Beno\^it \inst{\ref{Neel}}
    \and  S.~Berta \inst{\ref{IRAMF}}
    \and  M.~B\'ethermin \inst{\ref{Strasbourg}}
    \and  L.~J.~Bing \inst{\ref{Sussex}}
    \and  A.~Bongiovanni \inst{\ref{IRAME}}
    \and  J.~Bounmy \inst{\ref{LPSC}}
    \and  O.~Bourrion \inst{\ref{LPSC}}
    \and  M.~Calvo \inst{\ref{Neel}}
    \and  A.~Catalano \inst{\ref{LPSC}}
    \and  D.~Ch\'erouvrier \inst{\ref{LPSC}}
    \and  M.~De~Petris \inst{\ref{Roma}}
    \and  S.~Doyle \inst{\ref{Cardiff}}
    \and  E.~F.~C.~Driessen \inst{\ref{IRAMF}}
    \and  G.~Ejlali \inst{\ref{Teheran}}
    \and  A.~Ferragamo \inst{\ref{Roma}}
    \and  A.~Gomez \inst{\ref{CAB}} 
    \and  J.~Goupy \inst{\ref{Neel}}
    \and  C.~Hanser \inst{\ref{LPSC}}
    \and  S.~Katsioli \inst{\ref{AthenObs}, \ref{AthenUniv}}
    \and  F.~K\'eruzor\'e \inst{\ref{Argonne}}
    \and  C.~Kramer \inst{\ref{IRAMF}}
    \and  B.~Ladjelate \inst{\ref{IRAME}} 
    \and  G.~Lagache \inst{\ref{LAM}}
    \and  S.~Leclercq \inst{\ref{IRAMF}}
    \and  J.-F.~Lestrade \inst{\ref{LERMA}}
    \and  J.~F.~Mac\'ias-P\'erez \inst{\ref{LPSC}}
    \and  S.~C.~Madden \inst{\ref{CEA}}
    \and  A.~Maury \inst{\ref{Barcelona1}, \ref{Barcelona2}}
    \and  F.~Mayet \inst{\ref{LPSC}}
    \and  A.~Monfardini \inst{\ref{Neel}}
    \and  A.~Moyer-Anin \inst{\ref{LPSC}}
    \and  M.~Mu\~noz-Echeverr\'ia \inst{\ref{IRAP}}
    \and  I.~Myserlis \inst{\ref{IRAME}}
    \and  R.~Neri \inst{\ref{IRAMF}}
    \and  A.~Paliwal \inst{\ref{Roma2}}
    \and  L.~Perotto \inst{\ref{LPSC}}
    \and  G.~Pisano \inst{\ref{Roma}}
    \and  V.~Rev\'eret \inst{\ref{CEA}}
    \and  A.~J.~Rigby \inst{\ref{Leeds}}
    \and  A.~Ritacco \inst{\ref{LPSC}}
    \and  H.~Roussel \inst{\ref{IAP}}
    \and  F.~Ruppin \inst{\ref{IP2I}}
    \and  M.~S\'anchez-Portal \inst{\ref{IRAME}}
    \and  S.~Savorgnano \inst{\ref{LPSC}}
    \and  K.~Schuster \inst{\ref{IRAMF}}
    \and  A.~Sievers \inst{\ref{IRAME}}
    \and  C.~Tucker \inst{\ref{Cardiff}}
    \and  R.~Zylka \inst{\ref{IRAMF}}
    }
    
  \institute{
    Universit\'e C\^ote d'Azur, Observatoire de la C\^ote d'Azur, CNRS, Laboratoire Lagrange, France 
    \label{OCA}
    \and
    School of Physics and Astronomy, Cardiff University, Queen’s Buildings, The Parade, Cardiff, CF24 3AA, UK 
    \label{Cardiff}
    \and
    Universit\'e Paris-Saclay, Universit\'e Paris Cit\'e, CEA, CNRS, AIM, 91191, Gif-sur-Yvette, France
    \label{CEA}
    \and
    Institut de Radioastronomie Millim\'etrique (IRAM), Avenida Divina Pastora 7, Local 20, E-18012, Granada, Spain
    \label{IRAME}     
    \and        
    Aix Marseille Univ, CNRS, CNES, LAM (Laboratoire d'Astrophysique de Marseille), Marseille, France
    \label{LAM}
    \and
    Institut N\'eel, CNRS, Universit\'e Grenoble Alpes, France
    \label{Neel}
    \and 
    Universit\'e de Strasbourg, CNRS, Observatoire astronomique de Strasbourg, UMR 7550, 67000 Strasbourg, France
    \label{Strasbourg}
    \and
    Astronomy Centre, Department of Physics and Astronomy, University of Sussex, Brighton BN1 9QH
    \label{Sussex}
    \and
    Institut de RadioAstronomie Millim\'etrique (IRAM), Grenoble, France
    \label{IRAMF}
    \and
     Univ. Grenoble Alpes, CNRS, Grenoble INP, LPSC-IN2P3, 53, avenue des Martyrs, 38000 Grenoble, France
    \label{LPSC}
    \and 
    Dipartimento di Fisica, Sapienza Universit\`a di Roma, Piazzale Aldo Moro 5, I-00185 Roma, Italy
    \label{Roma}
    \and
    Univ. Grenoble Alpes, CNRS, IPAG, 38000 Grenoble, France 
    \label{IPAG}
    \and
    Institute for Research in Fundamental Sciences (IPM), School of Astronomy, Tehran, Iran
    \label{Teheran}
    \and
    Centro de Astrobiolog\'ia (CSIC-INTA), Torrej\'on de Ardoz, 28850 Madrid, Spain
    \label{CAB}
    \and
    National Observatory of Athens, Institute for Astronomy, Astrophysics, Space Applications and Remote Sensing, Ioannou Metaxa
    and Vasileos Pavlou GR-15236, Athens, Greece
    \label{AthenObs}
    \and
    Department of Astrophysics, Astronomy \& Mechanics, Faculty of Physics, University of Athens, Panepistimiopolis, GR-15784
    Zografos, Athens, Greece
    \label{AthenUniv}
    \and
    High Energy Physics Division, Argonne National Laboratory, 9700 South Cass Avenue, Lemont, IL 60439, USA
    \label{Argonne}
    \and  
    LERMA, Observatoire de Paris, PSL Research University, CNRS, Sorbonne Universit\'e, UPMC, 75014 Paris, France  
    \label{LERMA}
    \and
    Institute of Space Sciences (ICE), CSIC, Campus UAB, Carrer de Can Magrans s/n, E-08193, Barcelona, Spain
    \label{Barcelona1}
    \and
    ICREA, Pg. Lluís Companys 23, Barcelona, Spain
    \label{Barcelona2}
    \and
    IRAP, CNRS, Université de Toulouse, CNES, UT3-UPS, (Toulouse), France 
    \label{IRAP}
    \and
    Dipartimento di Fisica, Universit\`a di Roma ‘Tor Vergata’, Via della Ricerca Scientifica 1, I-00133 Roma, Italy 
    \label{Roma2}
    \and
    School of Physics and Astronomy, University of Leeds, Leeds LS2 9JT, UK
    \label{Leeds}
    \and    
    Institut d'Astrophysique de Paris, CNRS (UMR7095), 98 bis boulevard Arago, 75014 Paris, France
    \label{IAP}
    \and
    University of Lyon, UCB Lyon 1, CNRS/IN2P3, IP2I, 69622 Villeurbanne, France
    \label{IP2I}
  }
   \date{Received 14 March 2025; accepted 17 June 2025}

\abstract{The NIKA2 Cosmological Legacy Survey (N2CLS) is a large programme using the NIKA2 dual-band camera on the IRAM 30\,m telescope. Its goal is to improve our understanding of the physics of distant Dusty Star Forming Galaxies (DSFGs) by carrying out deep surveys of two fields, GOODS-North and COSMOS. This work is focussed on GOODS-North, which was observed for 78.2 hours, simultaneously at 1.2 and 2\,mm, with a field of view of $\sim$240\,arcmin$^2$. With such a deep integration, we were able to measure, for the first time, the confusion noise limits at the 30\,m telescope using the best sampled $\sim 62$\,arcmin$^2$ and masking sources with a flux greater than 0.54 or 0.17\,mJy at 1.2 or 2\,mm, respectively. We found a confusion noise of $139.1^{+ 15.9}_{- 19.2}\pm11.9$\,$\mu$Jy/beam at 1.2\,mm and $38.6^{+  9.6}_{- 13.1} \pm3.7$\,$\mu$Jy/beam at 2\,mm (the first uncertainty is statistical, the second is the cosmic variance). In this region, this corresponds to half the instrumental noise. To derive these estimates, we  devised a novel estimator, referred to as the cross variance, which also enabled us to estimate the correlated confusion noise between the two bands. Thus, we obtained a result of $49.6^{+ 15.9}_{- 24.8}\pm 6.4$\,$\mu$Jy/beam. These values are consistent with the state of the art Simulated Infrared Dusty Extragalactic Sky (SIDES) model.} 
 

\titlerunning{Confusion at the 30\,m telescope}
\maketitle

\section{Introduction}

In the far-infrared (FIR) to radio domains, single-dish surveys face a fundamental limitation in sensitivity and photometric accuracy due to fluctuations in the background surface brightness. In surveys of so called cosmological fields (i.e. free from Galactic contamination, the dominant source of background fluctuations is unresolved extragalactic point sources, which contribute to confusion noise in both the radio \citep{Condon1974} and IR domains \citep{Kiss+01, Lagache+03, Dole+04}. This confusion noise establishes the fundamental sensitivity limit for single-dish surveys, particularly for blind detection and accurate flux measurements of point sources.

The angular resolution of a single-dish telescope directly depends on the observed wavelength, limiting its ability to resolve individual sources at longer wavelengths. For example, the \textit{Spitzer} Space Telescope’s 85\,cm mirror \citep{Werner2004} resolved over 80\% of the galaxies contributing to the total emission at 24\,$\mu$m \citep{Papovich2004}, but only 23\% at 70\,$\mu$m and 7\% at 160\,$\mu$m \citep{Dole+04}. Similarly, the \textit{Herschel} Space Observatory \citep{Pilbratt2010}, with its larger 3.5\,m mirror, was able to resolve 58\% and 74\% of the sources at 100\,$\mu$m and 160\,$\mu$m, respectively, but only 25\% at 250\,$\mu$m \citep{Berta2011, Oliver2010}. To overcome these resolution limitations, innovative techniques have been developed, such as stacking analyses of known galaxy populations \citep{Dole2006, Bethermin2012b, Viero2013b} and analyses based on the probability of deflection to study unresolved source fluctuations \citep{Glenn2010, Berta2011, Bethermin2012b, Viero2013}. Additionally, studying the anisotropies of the cosmic infrared background (CIB) offers valuable insights into large-scale distribution and clustering of dusty star-forming galaxies, as well as the properties of the CIB itself \citep{Lagache2000, Amblard2011, Planck_CIB2011, Viero2013, Planck_CIB2013}. This helps achieve an effective approach to constrain the evolution of galaxy populations and cosmic star formation processes.

Following the initial theoretical studies of confusion noise in the radio domain \citep{Condon1974}, its first empirical assessment was made in a 4\,GHz radio survey using a half-difference technique to separate instrumental noise from observed noise \citep{Ledden1980}. This methodology has been extended to the FIR and submillimetre (submm) range, where confusion levels attributed to extragalactic sources and galactic cirrus fluctuations have been predicted \citep{Helou1990, Hacking1991, Gautier+92}. In the FIR domain, confusion noise was measured with ISOPHOT data \citep{Kiss+01}, which revealed fluctuations from extragalactic sources. Confusion noise, which represents the ultimate sensitivity limit for source detection, has become a critical parameter for assessing the performance of extragalactic deep fields. For the \textit{Spitzer} Space Telescope, studies of confusion noise were undertaken to optimise deep IR surveys, particularly in regions with high source densities \citep{Kiss+05}. Observations from the MIPS instrument at 24\,$\mu$m quantified confusion limits imposed by the high density of faint unresolved sources \citep{Dole+03}, influencing the design of deep field surveys \citep{Dole+04}. Measurements at longer wavelengths, such as 70\,$\mu$m and 160\,$\mu$m, revealed the combined role of extragalactic fluctuations and Galactic cirrus in setting confusion limits, providing a means for reliable source detection amidst the noise background \citep{Frayer2006b, Papovich2004}.

In the submm range, confusion noise also determines the depth of the survey. This was first addressed by \citet{Blain1998}, who highlighted the impact of confusion noise from faint, dusty galaxies on single-dish observations. The problem was further explored with SCUBA at the JCMT, where confusion from unresolved sources set the practical sensitivity limits for wide-field surveys \citep{Hughes1998, Barger1998, Eales1999}. More recently, observations with SCUBA-2 refined the understanding of submm confusion, particularly at 850\,$\mu$m. \citet{Geach2017} measured the confusion noise at 0.42\,mJy, highlighting the critical role of clustering in shaping the background fluctuations and limiting source extraction. The \textit{Herschel} Multi-tiered Extragalactic Survey (HerMES) provided a detailed assessment of confusion noise using SPIRE, quantifying detection limits at 250, 350, and 500\,$\mu$m \citep{Nguyen2010}. Follow-up analyses of \textit{Herschel} data deepened our understanding of confusion noise contributions from clustered dusty galaxies and Galactic cirrus, aiding in the optimisation of survey strategies and robust source extraction \citep{Lagache+03, Negrello+04, Nguyen2010}. These studies have proven instrumental in improving our ability to characterise the faint submm galaxy population, which dominates the cosmic IR background.

Different methods are commonly used to estimate the level of confusion noise, especially in the radio, FIR, and submm bands. One category of methods is based on galaxy number count models. A widely used approach calculates the number of sources per beam based on the source count distribution, $dN/dS$ \citep{Dole2003}. These methods estimate the confusion noise as the flux density variance induced by unresolved sources within the telescope beam. The flux cut-off threshold can be defined by a photometric criterion, such as a signal-to-noise ratio (S/N) threshold, or a source density criterion that assumes a uniform spatial distribution of sources \citep{Takeuchi2001}. While effective in sparse source environments, this approach becomes less reliable in clustered fields where source correlations amplify background fluctuations, rendering the uniformity assumption invalid.

Another type of method is data-driven and derives the confusion noise directly from the map noise measurements. In this approach, the confusion noise ($\sigma_c$) is derived as the asymptotic limit of the map noise ($\sigma_m$) as a function of observation time. The relation $\sigma_m^2 = \sigma_i^2 + \sigma_c^2$ is used, where $\sigma_i$ represents the instrumental noise, which decreases with time as $t^{-0.5}$ \citep{Nguyen2010,Geach2017}. This technique has the advantage that it does not rely on assumptions about the underlying source distribution. However, it is sensitive to systematic effects in the instrument or data reduction processes that may not diminish with longer observing times, potentially biasing the estimated confusion limit.

These methods highlight a trade-off between model-based and empirical approaches. While number count models make use of well-understood theoretical frameworks, they falter in complex fields with clustered sources. Conversely, empirical methods offer robustness to such clustering but are vulnerable to systematic errors, highlighting the need for careful cross-validation between methods in confusion-limited surveys.

NIKA2 is the current IRAM camera for measuring the millimetre continuum emission of astrophysical objects at the 30\,m telescope (Pico Veleta, Spain). It is described in detail, together with its performance in \cite{perotto20}. In short, it is a dual-band camera whose detectors are kinetic inductance detectors (KID) \citep{Monfardini2010}. The bands, which are observed simultaneously, are centred at 260 and 150\,GHz (1.2 and 2\,mm, respectively). There are three focal planes (arrays of detectors), two operating at 260\,GHz and consisting of 1,140 KIDs each, as well as one operating at 150\,GHz with 616 KIDs. About 90 and 84\% of these detectors are effectively used for observations. The photometric bandwidths are 49 and 40\,GHz. The main beams have FWHMs of 11.1 and 17.6\,arcsec  (at 1.2 and 2\,mm), and the field of view (FOV) is of 6.5\,arcmin in diameter. This large field of view combined with point source sensitivities of 30 and 9 mJy.s$^{1/2}$ provides mapping speeds of 111 and 1388\,arcmin$^2$.mJy$^{-2}$.h$^{-1}$ at 1.2 and 2\,mm respectively, an order of magnitude better than the previous generation photometric instruments at the 30\,m \citep{perotto20}. Although it is not used in this work, linear polarisation measurements can be made with the 1.2\,mm channel. NIKA2 has been available to the community since 2017. In return for its efforts to design, fund, install, and characterise the instrument, the NIKA2 collaboration\footnote{\url{https://nika2.osug.fr/-Collaborators-}} has been granted 1300\,h of guaranteed time of observation that have been divided into five large programmes. This paper is one of those presenting the results of the NIKA2 Cosmological Legacy Survey (N2CLS), which consists of a deep integration of two extragalactic fields with many available ancillary data: \gn\ and COSMOS.

This paper is organised as follows. In Sect.~\ref{se:gn}, we present the \gn\ field and our observations. In Sect.~\ref{se:da}, we describe the data processing that has been devised to produce the scientific maps. Sect.~\ref{se:measuring_confusion} presents the confusion noise estimator that we have built, the obtained measurements and their validation. In Sect.~\ref{se:confusion_sources}, we discuss the physical interpretation of the measured confusion. We present our summary and conclusions in Sect.~\ref{se:ccl}.

\section{Observing the \gn\ field}
\label{se:gn}

\gn\ is one of the two fields of the NIKA2 Large programme "NIKA2 Cosmological Legacy Survey" (N2CLS). An introduction to this IRAM 30\,m Large Programme and a first analysis of its observations in terms of deep cosmological number counts have been proposed in \citet{bing}. For the sake of clarity, we recall here the main features of our observations that are relevant to this work.

From October 2017 to March 2021, across 12~NIKA2 runs (i.e. one or two consecutive weeks of observations), we have observed \gn\ 749 times. In the following, we will refer to each of these observations as a 'scan'. The observed region is a rectangle of $12'\times 6.3'$ centred on RA=12:36:55.03 and Dec=62:14:37.59. Each of these raster scans is a collection of 'subscans' that encompass a straight trajectory of the telescope, that are parallel to each other and that are separated by $20''$. A scan has its subscans oriented at 40 or -50$^\circ$ in (RA, Dec) so that we cover exactly the same region each time, but with alternating orthogonal directions. This was decided in anticipation of robustness tests (see Sect.~\ref{se:results}). The scanning strategy on \gn\ was designed to reach sensitivities comparable to the predicted confusion at 1.2 and 2\,mm, respectively, estimated from simulations to be about 140 and 40\,$\mu$Jy/beam \citep{2017A&A...607A..89B}, in less than 100\,h of observations. Each scan lasted about 7 and 8\,mn.

During these 4.5 years of observations, some scans have suffered from poorer atmospheric conditions or instrumental instabilities. Thus, we  rejected the ones that proved to be more of a liability than an asset. Our main criterion was the final average noise of an individual scan map compared to that of all the other scans. For this work, to reduce the noise in the data, we used an improved version of the pipeline described in \citet{perotto20}, which has been used in previous papers published by our collaboration\footnote{\url{https://nika2.osug.fr/-Publications-} and \url{https://publicwiki.iram.es/Continuum/NIKA2/Main}} (see Sect.~\ref{se:data_model}). Although it is similar in its basic principles as far as noise reduction is concerned, this method also differs from the one used in \citet{bing}. Indeed, the need for extensive end-to-end simulations dedicated to this work was more conveniently achieved with the NIKA2 collaboration pipeline. Our methods have different exclusion criteria when it comes to selecting scans. In this work, we ended up using 668 scans, both at 1.2 and 2\,mm, thereby giving 78.2\,h of observation on the field. As for the opacities, the mean values are 0.25 and 0.15, with minima of 0.1 and 0.07, and maxima of 0.5 and 0.3 at 1.2 and 2\,mm, respectively.

\section{Data reduction}
\label{se:da}

In this section, we present our data reduction pipeline. We  retained the \textit{Planck} idiom and refer indifferently to timelines or time-ordered information (TOI) for the data streams recorded by the detectors over time.

\subsection{From raw to calibrated data}
\label{se:data_model}

Most of the low-level processing that goes from the raw data to the calibrated data is the same as that described in detail in \cite{perotto20}. Here, we present  only the modifications that are relevant to this work. In particular, we have used a novel method to monitor the KID resonance frequency shifts and convert them into a TOI that is proportional to the incident flux. This method, described in detail in Appendix~\ref{app:cf}, improves on the previous one in terms of the numerical integration, which improves the residuals from the atmosphere and low frequency electronic noise. Our data model has also changed. In \citet{perotto20},  for each detector, we estimated a single composite mode of atmosphere and electronic noise from the other detectors that are most correlated with it. This composite mode was computed while taking into account a bright source mask and both can be improved by an iterative scheme. While this method has proven effective for all the other science objectives of NIKA2, the level of integration is so much deeper in this work that it proved insufficient and led us to explicitly separate the atmosphere and several different 'modes' of the sky noise and the electronic noise. We explored several approaches based on more sophisticated instrument models. The main idea was to use the distribution of detectors per electronic box and their analogue sub-bandpasses to derive educated models of the noise modes. Although promising, this idea did not meet our expectations. The output maps were indeed very clean, but the transfer function (see Sect.~\ref{se:transfer_function}) was smaller than 50\%. This led to too much uncertainty in the actual blind detection of the bright sources to be masked and to a large and more uncertain correction in the final quoted values of the confusion.

In the end, we limited ourselves to one mode per acquisition electronics box. This mode, however, is not built up from the KID TOIs, but from measures of the same feedlines where there are no KID resonances. This is the analog of dark or blind detectors in other instruments and we refer to them as 'tones'. There are approximately 140 such measures available for each of arrays 1 and 3 (both at 1.2\,mm) and 18 for array 2 (at 2\,mm). This relatively low number is a limitation for our 2\,mm channel. The same circle method is used for these off-resonance tones as for the standard KIDs to derive a frequency variation. Empirically, we have found that the signal that is orthogonal to the circle frequency variation (see Appendix~\ref{app:cf}) is not useful for the decorrelation, so we used only the tangential on-circle variations, in the same way as for KIDs. Therefore, we here modelled the TOI of detector $k$ as

\begin{eqnarray}
m^k_t &=& \gamma^k P^k_{tp}S_p + \alpha_0^k A_t^{LF} + \alpha_1^k A_t^{HF} + \alpha_2^k\frac{\partial{A_t}}{\partial t} + \alpha_3^k A_t^2 \nonumber \\
&&+\alpha_4^k el_t +  \alpha_5^k+\alpha_6^k\,t+
\sum_{b=1}^{N_{b}}\beta^k_b E^b_t\nonumber\\
&& + \sum_{i=1}^{N_T}(\eta^k_i\cos(2\pi \nu_s\,i\,t) + \xi^k_i\sin(2\pi \nu_s\,i\,t)) + n^k_t\,.
\label{eq:data_model}
\end{eqnarray}

In Eq.~(\ref{eq:data_model}), $t$ is time, $P_{tp}$ is the pointing matrix, and $S_p$ is the sky brightness map in pixel, $p$. Here, $A_t$ is  the atmosphere as seen by all the KIDs, decomposed into its low frequency part: $A_t^{LF}$ below 0.5~Hz and its high frequency part: $A_t^{HF}=A_t-A_t^{LF}$. To the first order, all KIDs see the same atmospheric signal, since its origin is in the near field of the telescope. Then, $el_t$ is the elevation of the telescope and $E^b_t$ is the series of noise components derived from off-resonance tones, which are partially correlated with a range of different detectors depending on their common readout electronic box, $b$. Then, $n^k_t$ is the individual KID Gaussian white noise. We also included the time derivative of the atmosphere, $\frac{\partial{A_t}}{\partial t}$, as a proxy for second-order variations of the atmospheric contribution across the FOV, for possible residual  non-linearity with the $A_t^2$ term and a time linear drift, $\alpha_5^k+\alpha_6^k\,t$. Finally, on a subscan basis, we also defined harmonic modes to filter out low-frequency residual noise. To filter short and long subscans equally, we set the effective number of trigonometric modes per subscan, so that the highest subtracted frequency was 0.2\,Hz. In practice, this added  $N_T=4$ modes (2 cosine and 2 sine functions) for each of the 10\,s subscans (6x12 arcmin case) or 6 modes for each of the 17\,s subscans (12x6 arcmin case). Here, $\gamma^k$ is the absolute calibration, known from specific  observations (see \cite{perotto20}) on Uranus and skydips, while the $\alpha^k$, $\beta^k$, $\eta^k$, and $\xi^k$ coefficients are determined during the processing, as described in the next subsection. The actual implementation of this model goes along with the map making presented in the next section.

\subsection{Noise reduction and map making}
\label{se:map_making}

In all of this work, we built maps at a resolution of 4\,arcsec, so that it is less than a third of the main beam FWHM of the instrument, as the standard proxy for Nyquist's criterion. Our data reduction scheme is then:

\begin{enumerate}
    \item For each time, $t$, the median of the simultaneous samples of all the KIDs in the same array is computed. The median ensures robustness against possible outliers due to possible bright sources or anomalies that are seen by only a few KIDs at the same time, compared to the $\sim1\,000$ KIDs of the array. We internally iterate once on the mode production by eliminating KIDs that do not correlate with this initial mode (e.g. due to a bad tuning) and take the average of all selected KID TOIs.
    \item This mean mode is dominated by the atmosphere and can therefore be used as an estimate of $A_t$. Its derivative $\partial A/\partial t$ is derived after a 0.5\,s smoothing.
    \item One noise mode per acquisition box $E^b_t$ is built with the average signal of the off-resonance tones of this box.
    \item Each KID timeline, $m^k_t$, is simultaneously linearly regressed against the atmosphere modes, the noise modes, and the trigonometric functions to estimate all the coupling coefficients $\alpha^k$, $\beta^k_i$, $\eta^k_j$, and $\xi^k_j$. This regression is performed on a subscan basis to better account for the variability of the KID coupling to the different electronic modes. We also note that all electronic box modes are used altogether in the fit for each KID, and not just the box mode of that KID. This is to cope with the mixing of the electronic modes when $A_t$ and $\partial A/\partial t$ are generated using all KIDs of the same array.
    \item A few percent of the KIDs show a poor fit to this model and are excluded at this stage.
    \item For each remaining KID, the estimated atmosphere and noise contributions are subtracted, leaving only $\gamma^k P^k_{tp}S_p + n^k_t$ in Eq.~(\ref{eq:data_model}).
    \item Once all the KID timelines from a scan have been processed in this way, they can be projected onto a map. To do this, we use a standard nearest grid point average and apply a weight to each data sample. This weight is the inverse of the variance of the KID timeline during the subscan to which the data sample belongs. For the record, we note a slight overall excess of variance during the first two seconds of each subscan and we down-weight the corresponding data samples accordingly.
    \item Low-level residuals can be seen in maps in Nasmyth coordinates, with a striping oriented in the same way as the electronic boxes. We decided to correct for this pattern with one template per array and per subscan.
\end{enumerate}

\begin{figure*}
    \centering
    \includegraphics[width=7cm]{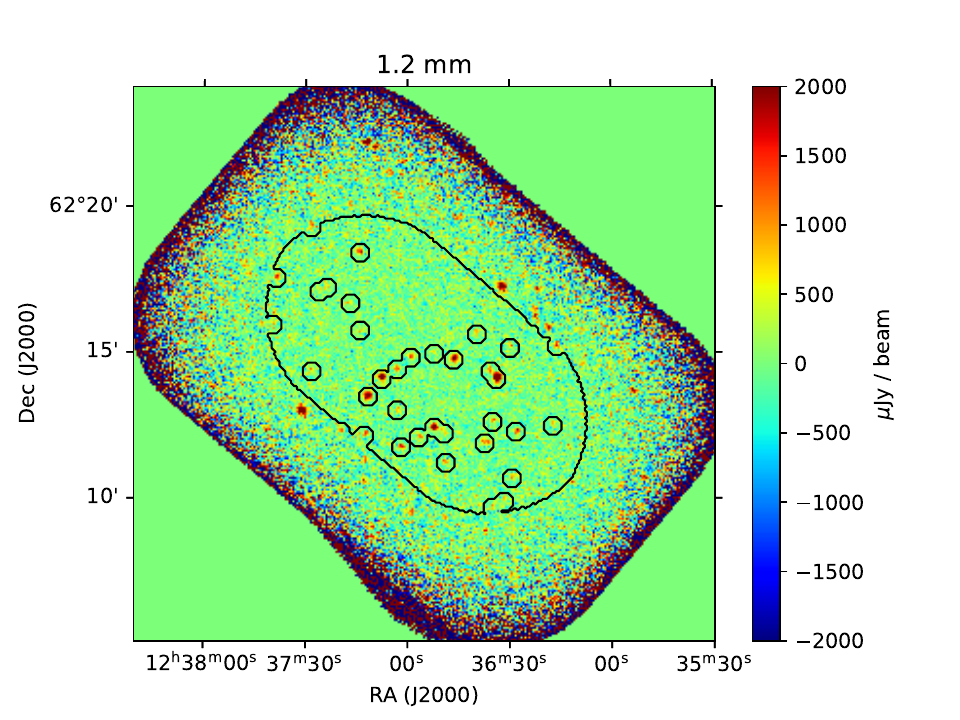}
    \includegraphics[width=7cm]{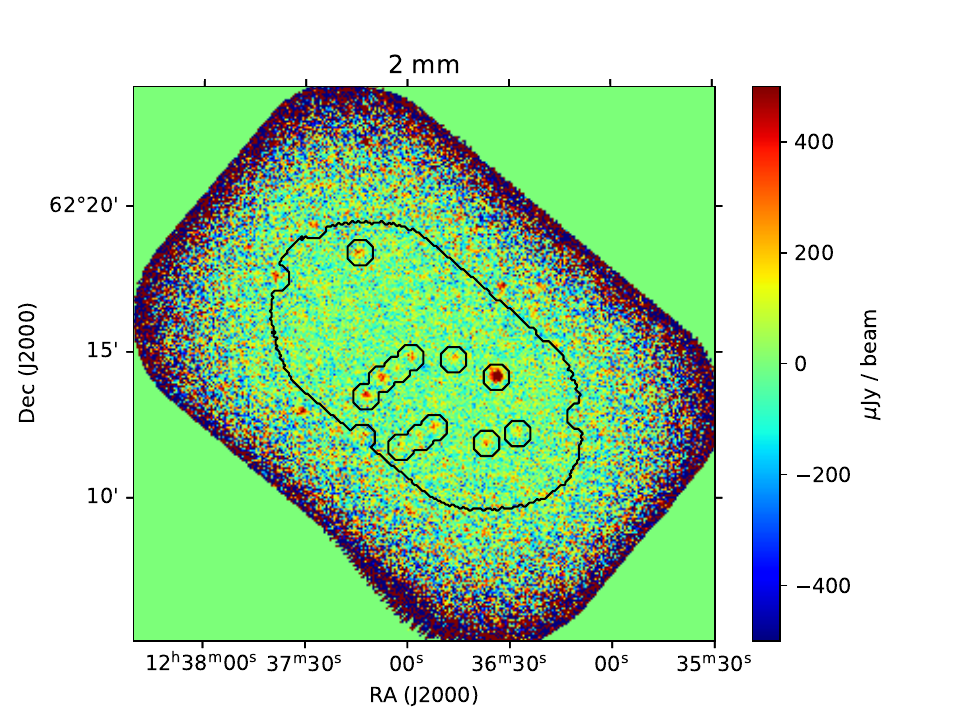}
    \caption{Maps of \gn\ used in this work (top 1.2\,mm, bottom 2\,mm). The black contours define the region (i.e. mask)  that we use to compute the confusion.}
    \label{fig:maps}
\end{figure*}

Each scan was reduced according to this scheme and we obtained a map per scan. We then co-added all these scan maps to produce the final map, using inverse noise per pixel weighting. This process is based on the implicit assumption that the signal is zero or at least negligible compared to the parasitic components. To improve the final map, we iterated this process. We performed a point source detection on the map and kept only the brightest ones that appear at a SNR greater than 10. We built a model map that is just the sum of these bright point sources modelled as constant-width 2D Gaussian. At the start of the next iteration, we de-projected this model map into point source TOIs, which we then subtracted from the KID TOIs. In this way, the final map was no longer affected by the filtering residuals of the bright sources. In practice, we proceeded more progressively. At each iteration, $iter$, we actually subtracted $(1-0.2^{iter})$ times the source model map built at the previous iteration to mitigate the effect of errors on this map at the very first iterations. This process was iterated until convergence was achieved, with residual peaks on the difference between successive iterations at less than $10\,\rm\mu Jy$. In practice, three processing iterations were sufficient, the first one assuming a vanishing point-source model map. Figure~\ref{fig:maps} presents our final maps, together with the contours of the region used to derive the confusion (cf.~Sect.~\ref{se:measuring_confusion}). The noise map was derived in two steps. As described in step 7 of our data processing, each sample was projected onto the final map with a known weight. Straightforward algebra allowed us to derive the total noise per pixel on the final map. However, we note that the distribution of the S/N per map pixel did not follow the expected normalised Gaussian, probably due to some residual correlation in the noise. We concluded that our TOI noise propagation underestimated the effective noise on the map. To correct for this, we needed to multiply our noise maps by typically 1.6 and 1.3 at 1.2 and 2\,mm, respectively.

\section{Measuring the confusion}
\label{se:measuring_confusion}

The confusion, $\sigma_c$, is defined as the square root of the variance of the measured brightness, $S$ (i.e. including the point spread function; PSF or beam of the instrument) when bright sources above a certain flux are masked out and in the limit of zero instrumental noise or (equivalently) infinite integration time \citep{Condon1974,2010A&A...518L...5N,2024A&A...692A..52B} as follows:

\begin{equation}
    \sigma_c^2 = \frac{1}{N}\sum_{p=1}^N (S_p-\bar{S})^2\, ,
    \label{eq:confusion_def}
\end{equation}

where $S_p$ is the signal at pixel, $p$, and $\bar{S}$ is the average brightness of the map. With this definition, the confusion depends on both the angular resolution of the instrument and the arbitrary flux cut of bright sources. In practice, maps are affected by the data processing and contaminated by instrumental noise so that Eq.~(\ref{eq:confusion_def}) cannot be readily implemented. In the next subsection, we present our estimator of $\sigma_c$. Our final results and a discussion on their robustness are summarised in Sect.~\ref{se:results}.

\subsection{The \xrms\ estimator}
\label{se:xrms}

To define the sky area over which we calculate the confusion, we have to make a trade-off between the size of the patch and the homogeneity of the noise. To this end, we chose to restrict to pixels with a number of samples greater than two times the median number of hits per pixel on the full map for both bands. We built on the results of \citet{bing} and masked sources that are confidently detected above a minimum S/N of 4.6 and correspond to fluxes greater than $\sim 0.54$ and $\sim 0.17$ mJy at 1.2 and 2\,mm, respectively. We took the locations of these sources and mask all pixels within a radius of 19~arcsec at 1.2\,mm and 27.8~arcsec at 2\,mm. This leaves us with 64.7 and 62.1 arcmin$^2$ at 1.2\,mm and 2\,mm (see Fig.~\ref{fig:maps}). The masks were independently derived at 1.2 and 2\,mm.

Implementing Eq.~(\ref{eq:confusion_def}) directly on the data leads to an estimate of $\hat{\sigma_c}^2$ that is biased. In fact, the measured value in each pixel, $A_p$, is the sum of the signal, $S_p$, and the noise $a_p$, so the two variances add up. Furthermore, the signal has been affected by the data processing. In our case, where we are interested in a single number, the effect of the processing (i.e. the transfer function) can be represented by a single factor, f. In the limit of zero noise on the maps, replacing $S_p$ by $A_p$ in Eq.~(\ref{eq:confusion_def}) gives $f^2\sigma_c^2$. The derivation of $f$ is detailed in Sect.~\ref{se:transfer_function} and is based on simulations. We went on to determine how we could estimate the variance of the noise, $\sigma_a^2$ . We  considered several possibilities.

First, we could use a null map produced by combining the 668 scans with alternating positive and negative weights. This would cancel out the astronomical signal, while leaving the noise properties unchanged. Applying Eq.~(\ref{eq:confusion_def}) to this map would directly give $\sigma^2_a$. In practice, the noise on the null map actually appears to be less than the noise on the map. Indeed, when we computed the histogram of the S/N, with the noise derived from the null map, its width is rather 1.2 than 1. We attribute this to residual noise in the data maps that is difficult to estimate precisely and that may bias our estimate of $\sigma_c$. Nevertheless, if we computed $\sigma_c = \sqrt{\sigma^2-\sigma^2_a}$ using this method, we would find $\sigma_c \simeq 145$ and 34\,$\mu$Jy/beam at 1.2 and 2\,mm, respectively. These values are in good agreement with our final results.

\begin{figure*}
    \centering
    \includegraphics[width=7cm]{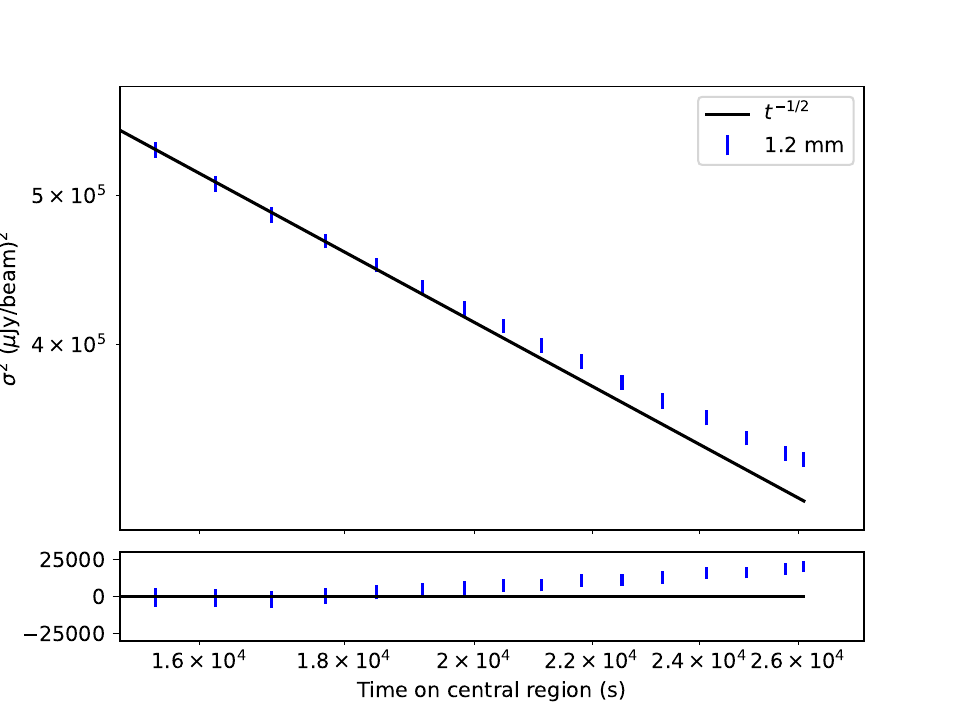}
    \includegraphics[width=7cm]{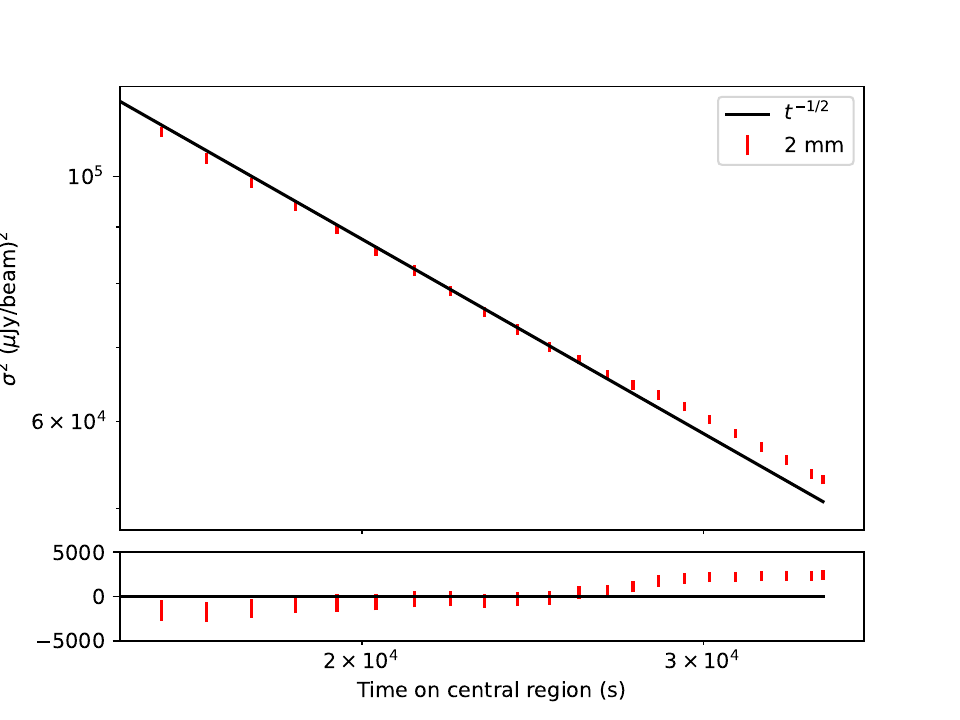}
    \caption{Variance of the N2CLS maps vs the out of atmosphere integration time on a region in the center equivalent to one NIKA2 field of view (6.5 arcmin diameter). To determine these data points, we simply take the variance of the unmasked pixels (see Fig.~\ref{fig:maps}) in the central area. For each NIKA2 band, the lower plot shows the residuals compared to the $t^{-1/2}$ instrumental noise integration that is fitted here for times of integration smaller than 20\,000\,s. The error bars are the square roots of the diagonal elements of the covariance matrix of the measures (Fig.~\ref{fig:rho_covar_matrix}), derived from 100 Monte-Carlo simulations of map-based realistic noise (per scan). There is a clear excess of signal at high integration time due to the confusion, but as explained in the main text, this plot is for illustrative purposes only and is not used to derive our final results.}
    \label{fig:sigma2_cumul}
\end{figure*}

Second, we could use an approach such as the one proposed in \citet{2010A&A...518L...5N} and monitor the measured noise as a function of the integration time. Unlike them, we did not have enough integration time to do this analysis pixel per pixel, but we could generalise their approach to a sub-area of our map, accumulating scans, calculating the variance of the map pixels as we went along, and then fit its decrease as a linear function of the inverse of the integration time. The noise variance would decrease as $1/t$ and the confusion would be the remaining constant term in the fit\footnote{To be more specific, this integration time is meant as out of the atmosphere. It is the actual wall clock observation time corrected for elevation and opacity, and as such, is different at 1.2 and 2\,mm. See  Sect. 10.2.1 in \cite{perotto20}.}. We  found that this does not solve the previous problem and comes with other difficulties, such as the determination of the  (strongly correlated) uncertainties of each measurement of $\sigma^2(t)$ and the determination of the out-of-atmosphere equivalent integration time. Both of these factors impact the derived value of the confusion and its associated uncertainty. The former can be addressed by Monte-Carlo simulations, but the latter is poorly defined in the context of a non-uniform survey and is affected by uncertainties in the opacity as well as the overall absolute calibration; this can vary between different runs, depending on the number and quality of absolute calibration scans that could be performed. In Fig.~\ref{fig:sigma2_cumul}, we still show the decrease of the variance of the map versus the integration time, but this is for illustrative purpose only. As such, the solid black line follows $t^{-1/2}$, which was fitted for integration times smaller than 20\,000\,s and it does not take the correlation between the error bars into account. For reference, completing this exercise yields $\sigma_c \simeq$~142 and 49\,$\mu$Jy/beam at 1.2 and 2\,mm, respectively. Given the actual significance of our detection (see below), the observed excess is real and dominated by the confusion and the agreement between this method and our final one is not surprising. However, the caveats we  mention here have prevented us from using this estimator to derive our final results.

To improve on all these aspects, we  developed a third approach based on the general idea of cross-correlations \citep[e.g.][]{2005MNRAS.358..833T}. Instead of combining all the scans into a single map, we combined them into a set of independent maps (1 for each of the 12 observation runs). This gives us 66 pairs of different maps of the same sky (excluding auto-correlations) and for each pair $(A_i,B_i)$, we can compute the cross-variance, as defined by:
\begin{equation}
\rho_i \equiv \frac{1}{f^2}\frac{1}{N}\sum_p (A^i_p-\bar{A}^i)(B^i_p-\bar{B}^i)\, ,
\label{eq:rho_ab_def}
\end{equation}
where $i$ is the pair index. Since the maps $A^i$ and $B^i$ are independent, the statistical expectation of $\rho_i$ for unmasked pixels is directly $\sigma^2_c$ (see eqs.~\ref{eq:rho_expect} and \ref{eq:f_and_rho}), without any bias. Indeed, since maps $A^i$ and $B^i$ were not coadded before computing the variance of the sum, their noise components did not add up in quadrature as in the previous two estimators. Finally, $\rho_i$ makes no assumption about the uniformity of the integration time per pixel and is therefore immune to inhomogeneities in the sky coverage. To make the best use of all the available data, we computed the cross-variance of all the possible pairs and determine their full covariance matrix, $X$ (see Appendix~\ref{app:equations} for details). With these two in hand, we were able to compute the likelihoods of the confusion at 1.2\,mm, 2\,mm, and even the cross-confusion $1.2\times 2$\,mm:
\begin{equation}
    \mathcal{L} \propto \exp{[-(\rho_i-\sigma_c^2)X_{ij}^{-1}(\rho_j-\sigma_c^2)]}\,.
\end{equation}

We relied on these likelihoods to derive final values and their associated confidence intervals. Our final results are discussed in Sect\,\ref{se:results}.

\subsection{Determining the transfer function}
\label{se:transfer_function}

To determine the transfer function factor, $f$, we relied on the state-of-the-art SIDES-UCHUU simulations \citep{2017A&A...607A..89B,2022A&A...667A.156B,2023A&A...670A..16G}. We applied the NIKA2 bandpass and point spread function to the SIDES-UCHUU cube to create a simulated sky, $sim_p$, as it would be observed by NIKA2. We then ran our exact scan strategy and KID selection over this map, for each scan, to build simulated signal timelines, $sim^k_t$, for each KID, $k$. These simulated timelines can then be added to our real data timelines. To be immune to the true astronomical signal in our data, before adding the simulated timeline, we multiplied our data timelines by -1 for every second pair of scans (to account for both scan orientations), so that the true signal is cancelled out in the final map, leaving only the simulated component. This is the best possible simulation to characterise our pipeline, as it deals with the true atmosphere and electronic noise rather than approximations. The magnitude of the simulated timelines is smaller than these two components by a factor of more than 1\,000, so it does not bias their derivation. We then reduced these composite timelines in exactly the same way as the data and project them onto a map, $M_p = sim_p^{processed} + N_p$, where $N_p$ is the noise. Basically, to determine $f$, we need to measure the rms of $sim_p^{processed}$ and divide it by the variance of the input noiseless simulated map, $sim_p$. A first way to proceed is to process the data timelines with exactly the same mask and alternating sign weights, but without adding $sim^k_t$. This gives $N_p$ which can be subtracted from $M_p$ to get $sim_p^{processed}$. A second way is to do the same whole process as on the data, namely making maps per runs, calculating their cross-variance\ and deriving the estimated final confusion as the value that maximises their likelihood. The ratio of this value to the standard deviation of the input map, $sim_p$, on the same area gives an estimate of $f$. As described in Sect.~\ref{se:gn}, our observations consist of two sets of scans with orthogonal directions and different subscan lengths. This results in slightly different effective filtering between these two types of scans, so we computed a transfer function, $f$, per direction and we corrected accordingly. We discuss this in more detail in Sect.~\ref{se:results}. We checked the robustness of our estimates of $f$ by varying the mask definition criteria (size and masked sources) and by scaling the SIDES simulations by factors ranging from 1 to 3. The scaling improves or degrades the S/N of the output signal, while remaining negligible compared to the true noise of the data. Finally, we used the $f$ factors derived from simulations scaled by a factor 3, as they provide a better dispersion due to their higher S/N. We  estimated $f$ on 27\footnote{The SIDES maps used in this work were more conveniently grouped by 9. As the agreement between the two derivations and the dispersion was satisfactory, there was no need to run further (rather long) end-to-end simulations.} simulations and the two methods agree on average by 1\%. The statistical uncertainty on the average of $f$ is also about 1\%, so we take this as our final uncertainty on $f$.

The dispersion of $\sigma_c$ as estimated from these 27 end-to-end simulations (including the detection of bright sources and masks) also provides an estimate of the cosmic variance associated with our measurement. We find about 12, 4, and 6\,$\mu$Jy/beam at 1.2, 2, and $1.2\times 2$\,mm, respectively (exact values are given in Table~\ref{tab:results}).

\subsection{Results}
\label{se:results}

\begin{figure*}
\centering
\includegraphics[width=5cm]{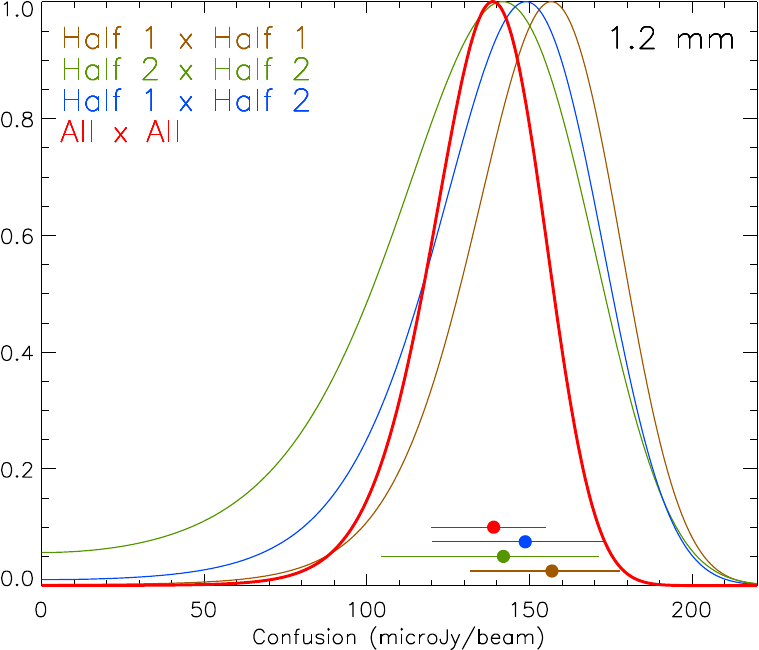}
\includegraphics[width=5cm]{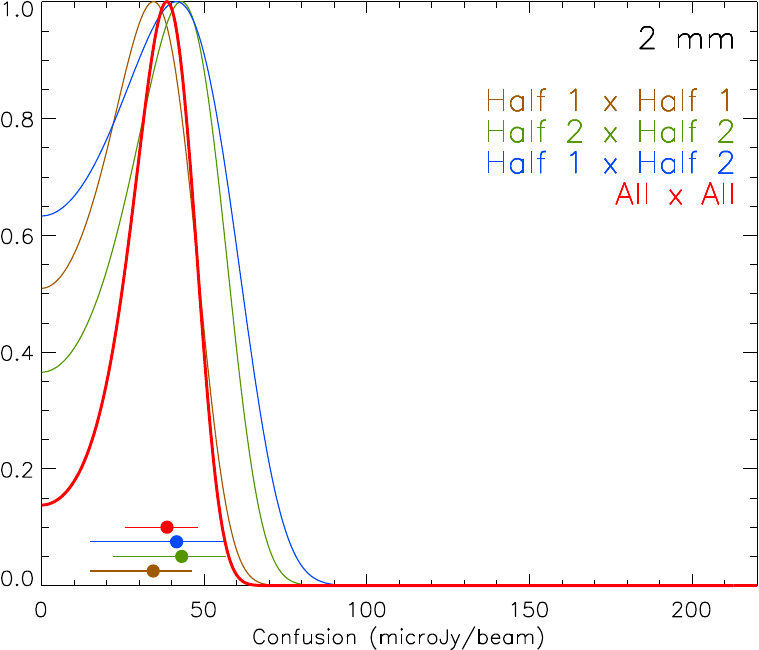}
\includegraphics[width=5cm]{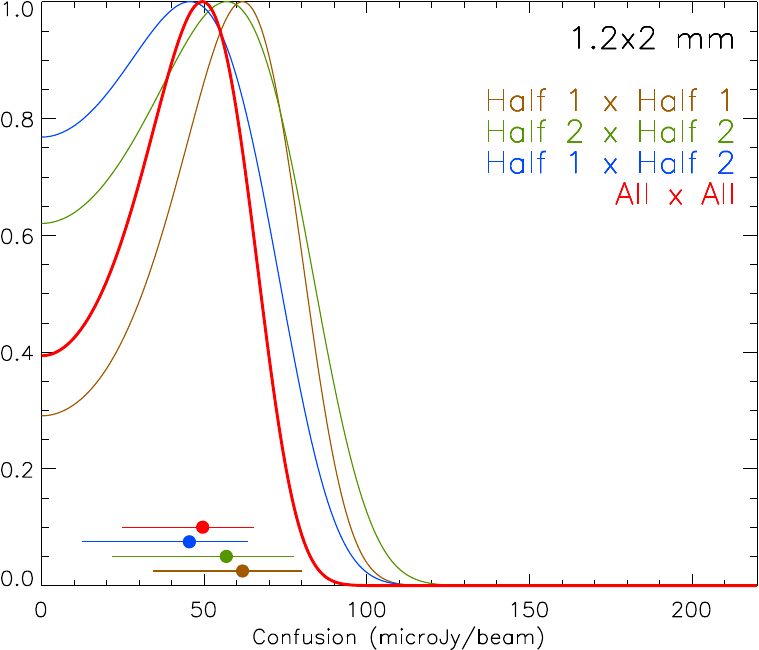}
\caption{Likelihoods (normalised to their maximum) of the confusion at 1\,mm, 2\,mm, and the cross-confusion 1$\times$2\,mm for different subsets of our data. The central points and their 1-$\sigma$ equivalent error bars are shown for convenience and have arbitrary ordinates. 'Half 1' refers to scans oriented at $40^\circ$. 'Half 2' refers to scans oriented at $-50^\circ$. 'All' refers to the total sum of the two.}
\label{fig:likelihoods}
\end{figure*}

The 12 observation runs give us 66 cross-variance\ measurements, which are given in Fig.~\ref{fig:all_rho} in the particular case of 1.2\,mm. The likelihoods of the confusion at both wavelengths are shown in Fig.~\ref{fig:likelihoods} as estimated on different subsets of our data. 'Half 1' denotes the ensemble of 334 of our scans that are oriented at 40$^\circ$, 'Half 2' is the 334 at $-50^\circ$. To check for consistency, we compute the confusion either exclusively on each half of the scans, on 'all' the scans combined or exclusively by crossing one half with the other. As mentioned in the previous section, we computed an effective transfer function, $f$, and corrected for it independently in each case. Indeed, this $f$ factor relates to the variance of the signal and is implicitly azimuthally symmetric, whereas our filtering is mostly per subscan and is therefore directional. This particular feature prevents us from deriving $f$ based on the combined directions taken from $f$ coming from each direction. As shown in Fig.~\ref{fig:likelihoods}, all estimates are compatible within 1\,$\sigma$. Using the values derived from all the scans (as they have the highest S/N), we obtained the confusion estimates shown in Table~\ref{tab:results}; namely, $139.1^{+ 15.9}_{- 19.2}\pm11.9$, $38.6^{+  9.6}_{- 13.1} \pm3.7$ and $49.6^{+ 15.9}_{- 24.8}\pm 6.4$\,$\mu$Jy/beam at 1.2, 2 and $1.2\times2$\,mm, respectively. The first set of error bars refers to the instrumental noise and the second  to the cosmic variance. These are clear measures at a significance of $7 \sigma$ at 1.2\,mm and $3 \sigma$ at 2\,mm, but only a $2 \sigma$ evidence at $1.2\times2$\,mm. An overall absolute calibration uncertainty has to be considered. It affects the quoted value and its statistical error bars, but not the cosmic variance. This uncertainty has three main components. Firstly, there is the statistical variability of the measurement under various observation conditions from the ground, which was determined to be 5\% during the NIKA2 commissioning. Secondly, there is uncertainty regarding the exact flux of our calibration source Uranus. As reported in  \citet{perotto20}, this uncertainty is about 5\%. Taking all these uncertainties into account and adopting a conservative approach, we estimated an overall uncertainty of 10\% in our absolute calibration.

To check the consistency of our final values, we used the SIDES simulations and the bright source masks derived from our end-to-end simulations and calculated the expected confusion levels. We find 156, 49, and 84 $\mu$Jy/beam at 1.2, 2, and $1.2\times 2$\,mm. If we compare this to our measures taken at face value, at 1.2 and 2\,mm, our results are compatible with SIDES at $1\sigma$ (stat), even without considering the uncertainty on absolute calibration or the cosmic variance. Our measurement of the $1.2\times 2$\,mm confusion is $1.4\sigma$ (stat) lower than the value expected from SIDES. Given the 10\% uncertainty on our absolute calibration, if we allow it to increase by this factor, our results at 1.2 and 2\,mm become $153\pm 21$ and $43\pm 14.  $  This shows an even better agreement with the prediction of SIDES. Our measurement at $1.2\times 2$\,mm becomes $55\pm28$\,$\mu$Jy/beam, which  is now $1.04\sigma$ (stat) of the 84\,$\mu$Jy/beam expected from SIDES. This last compatibility can be further improved if we assume that our field has a low intrinsic confusion within one cosmic standard deviation (Table~\ref{tab:results}). This would actually also improve the agreement at 2\,mm, but slightly degrade that at 1.2\,mm, although keeping it within $1\sigma$ (stat). Overall, this comparison shows a good agreement between our results and the predictions of SIDES at 1.2 and 2\,mm, but a marginally lower than expected value at $1.2\times 2$\,mm.

The consistency of our results can be assessed in another way. Using Eq.~(\ref{eq:confusion_no_clustering}), we can see that $\sigma_{c_2} = r \sigma_{c_1}\theta_2/\theta_{1.2}$ and $\sigma_{c_{1\times2}} = (2r)^{1/2}\sigma_{c_1}\theta_2/(\theta_1^2+\theta_2^2)^{1/2}$, where $\theta$ is the instrumental beam FWHM and $r$ is a constant that can be derived from the SIDES confusions and is 0.21. Applying this to the measured value of the confusion at 1.2\,mm and considering only the statistical uncertainty, we would expect $\sigma_{c_2}\simeq 44 \pm 6$ and $\sigma_{c_{1\times2}} = 75 \pm 10\,\mu$Jy/beam. Both of these values are compatible with our measures and their associated error bars, although, again, the measured value at $1.2\times2$\,mm appears slightly lower than the predicted one.

Finally, we compare these estimates with the instrumental noise in the final maps. On the central region of our maps equivalent to one NIKA2 field of view (a disk of 6.5\,arcmin diameter) and still masking bright sources, we find total variances of map pixels of about $2.8\times 10^5$ and $4.5\times 10^4$\,($\mu$Jy/beam)$^2$. Subtracting our estimates of confusion (139 and 39\,$\mu$Jy/beam) according to Eq.~(\ref{eq:instr_noise_per_beam}), we can derive the residual noise contribution per beam to be 218 and 60\,$\mu$Jy/beam at 1.2 and 2\,mm, respectively. It means that the measured confusions are 0.64 and 0.65 times the residual noise in our final maps. These noise estimates agree within 25\% with those obtained by \cite{bing} (170 and 48\,$\mu$Jy/beam) with a data reduction optimised for point source detection; thus, it filters  out more of the undetected sources that generate the confusion.

This was achieved after about 80~hours of observation of about 64\,arcmin$^2$, that is, approximately 11~hours per field of view of 33.2\,arcmin$^2$. These numbers provide with another cross-check when we use them to derive an approximate value of the instrumental sensitivity, also known as the Noise Equivalent Flux Density (NEFD). A thorough assessment of this sensitivity and its uncertainty with these data is beyond the scope of this paper. Still we note that taking the average encountered opacities and elevations (Sect.~\ref{se:gn}), we obtain effective out-of-atmosphere NEFDs of 32 and 10\,mJy/beam at 1.2 and 2\,mm, respectively. Although these values are only approximate, they are in good agreement with those presented in \cite{perotto20} and comfort our results.

\section{Sources responsible for the confusion}
\label{se:confusion_sources}
The confusion is caused by all galaxies with flux below the detection limit $S_{\rm lim}$. However, depending on the exact shape of the number counts, the confusion noise can be dominated by sources with different flux regimes. In this section, we use the SIDES simulation \citep{Bethermin2017} to predict the contribution of the different flux densities to the confusion, and compare it with the contribution to the CIB, expressed as
\begin{eqnarray}
B &=& \int_0^\infty S\frac{dN}{dS} dS\,, \nonumber \\
&=&\int_0^\infty S^2\frac{dN}{dS} \ln(10)\,  d\log_{10}(S)\,,
\end{eqnarray}

where $\frac{dN}{dS}$ is the differential number counts. The factor $S^2\frac{dN}{dS}$ thus directly gives the relative contribution of a logarithmic flux density interval to the background.

In the absence of clustering, the confusion noise is expressed as\begin{equation}
\sigma_c = \int\int b^2 d \Omega \int_0^{S_{\rm lim}} S^3\frac{dN}{dS} \ln(10) \, d\log_{10}(S),
\label{eq:confusion_no_clustering}
\end{equation}
where $b$ is the beam function. Similar to the background case, the term $S^3\frac{dN}{dS}$ gives the contribution of the logarithmic flux density intervals to the fluctuations. As shown in \citet{Bethermin2017}, the clustering tends to broaden the pixel flux distribution and thus increases the confusion. However, even for \textit{Herschel}/SPIRE at 500\,$\mu$m with its 36\,arcsec beam, the effect is of the order of 15\,\%. We can therefore reasonably neglect it in the case of NIKA2, which has a beam area four times smaller. 

\begin{figure*}
\centering
\includegraphics[width=6cm]{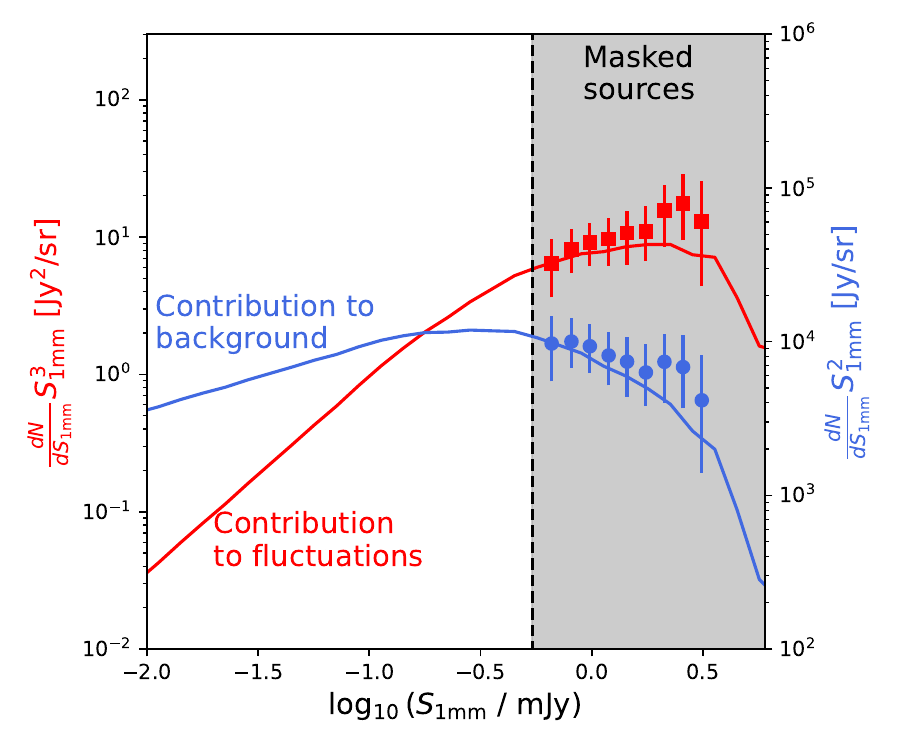}
\includegraphics[width=6cm]{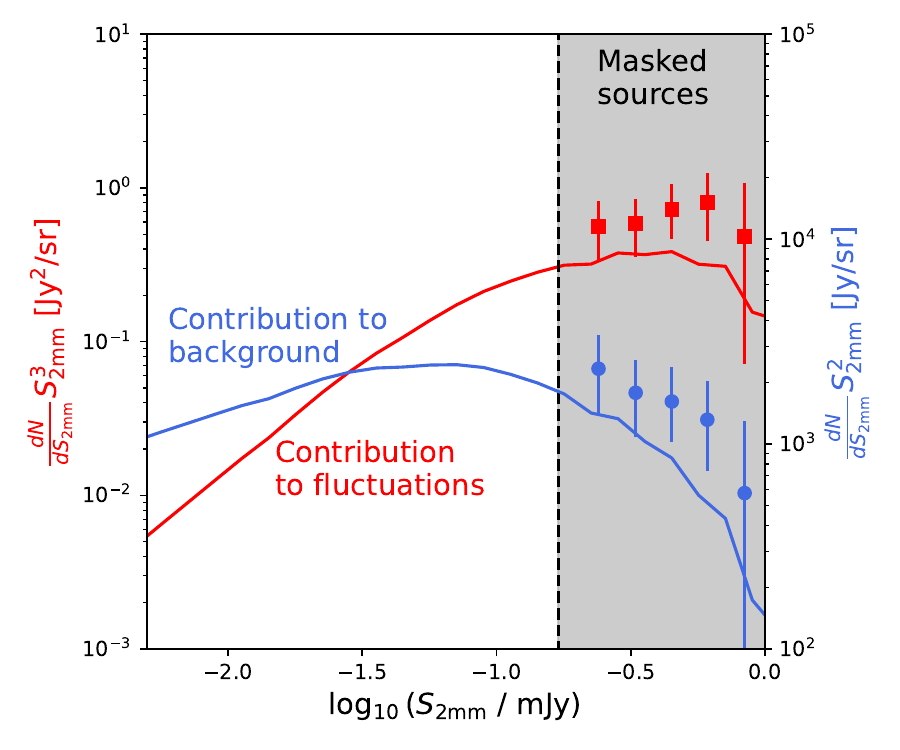}
\caption{Relative contribution to the background (blue curve) and to the fluctuations causing the confusion (red curve) as a function of the flux density derived with the SIDES simulation at 1.2\,mm (top) and 2\,mm (bottom) (see Sect.\,\ref{se:confusion_sources}). The same quantities derived from the source counts measurements in GOODS-N by \citet{bing} are shown as blue circles and red squares, respectively. The difference between the prediction of galaxy counts from SIDES and the source counts from N2CLS is well understood and caused by the blending of galaxies into the NIKA2 beam (see the analysis in \citealt{bing}). The grey area shows the flux density regime where the sources are masked and do not contribute to the confusion.}
\label{fig:confusion_contrib}
\end{figure*}

In Fig.\,\ref{fig:confusion_contrib}, we show the relative contribution to the background (blue) and the confusion (red) as a function of the flux density. Only sources below the detection limit contribute to the confusion and the grey area indicates the region where they are masked and therefore do not contribute. The curves are calculated from the number counts derived from the SIDES simulation. At both 1.2\,mm and 2\,mm, the maximum of the red curve is in the masked regime and decreases strongly with decreasing flux density. The sources just below the detection limit are therefore the ones that produce most of the confusion we measure. 
The contribution to the background peaks at fainter flux densities, which are below the detection limit. However, we note that at 1.2\,mm, this peak is just below the detection limit. At this wavelength, the flux density regime that dominates both the background and the confusion is therefore the same. 

The agreement of the confusion noise from SIDES and N2CLS is an important validation of SIDES in the flux regime corresponding to the peak of the contribution to the background and, thus, to the obscured star formation budget. At 2\,mm, the peak is below the detection limit by a factor of $\sim$3. Therefore, we are mainly probing galaxies that are slightly brighter than those that make up the bulk of the background.

\begin{table*}[htbp]
    \centering
\caption{Confusion noise estimates and uncertainty budget.}
\begin{tabular}{|l|c|c|c|}
\hline
& 1.2\, mm & 2\,mm & $1.2 \times 2$\,mm\\
\hline
$\sigma_X$ & $76.0^{+ 8.7}_{-10.5}$ & $20.4^{+ 5.1}_{- 6.9}$ & $23.4^{+ 7.5}_{11.7}$ \\
Transfer function $f$ & $0.60 \pm 0.01$ & $0.51 \pm 0.01$ & $0.48 \pm 0.02$\\
Abs. calibration & $1 \pm 0.1$ & $1 \pm 0.1$ & $1 \pm 0.1$ \\
Cosmic Variance & $11.9 \pm 1.6$ & $ 3.7 \pm 0.5$ & $ 6.4 \pm 0.9$ \\
DSFG color correction & 1.09 & 0.96 & 1.02 \\
\hline
Final Confusion & $(1.0\pm 0.1)\times(139.1^{+ 15.9}_{- 19.2}) \pm  11.9$ & $(1\pm 0.1)\times( 38.6^{+  9.6}_{- 13.1}) \pm   3.7$ & $(1\pm 0.1)\times( 49.6^{+ 15.9}_{- 24.8}) \pm   6.4$ \\
\hline
\end{tabular}

\tablefoot{$\sigma_X$ is the measured cross-variance before any correction. The transfer function factor $f$ (Eq.~(\ref{eq:f_and_rho})) is dimensionless, all other quoted values in this table are in $\mu$Jy/beam, where 'beam' refers to a fiducial Gaussian of 12.5 and 18\,arcsec FWHM at 1.2 and 2\,mm. In the case of $1.2\times2$\,mm cross-confusion, it refers to the square root of the product of the two beams. Statistical uncertainties on the confusion come from the limits of the 68\% confidence interval of the likelihoods. Uncertainties on the transfer function are given for reference but are negligible, and the estimate of the cosmic variance comes from 27 SIDES-UCHUU simulations that have been processed end-to-end. The final estimate of the confusion and its statistical error bars are from those of $\sigma_X$ corrected by the transfer function. The quoted value and its statistical uncertainties (but not the cosmic variance) are in total uncertain by about 10\% due to the uncertainty in the absolute calibration of the instrument (see Sect.~\ref{se:results}). Note that the NIKA2 absolute calibration is derived from laboratory measurements and observations of Uranus. We  applied the color correction factors given in the table to the final results, assuming a DSFG spectrum of $\sim \nu^{3.5}$.}
    \label{tab:results}
\end{table*}

\section{Conclusion}
\label{se:ccl}

From October 2017 to March 2021, we observed the \gn\ field. We were able to accumulate 78.2\,h of observations on about 240\,arcmin$^2$. By masking the sources that we  confidently detected in a previous work \citep{bing}, whose flux is greater than 0.54 or 0.17\,mJy at 1.2 and 2\,mm, and by restricting ourselves to the 62\,arcmin$^2$ of our best sampled observations, we were able to derive the first estimates of the confusion ever obtained at the IRAM 30\,m telescope. We measured $139.1^{+ 15.9}_{- 19.2}\pm11.9$\,$\mu$Jy/beam at 1.2\,mm, where the first set of error bars is due to noise and the second one due to the cosmic variance. At 2\,mm, we measured $38.6^{+  9.6}_{- 13.1} \pm3.7$\,$\mu$Jy/beam. To derive these estimates, we  developed a new estimator, referred to as the cross-variance. This estimate is less sensitive to residual noise on the maps and also allows us to measure the contribution to the confusion that is correlated between 1.2 and 2\,mm. In this case of cross-band confusion, we only found evidence at the level of a $2\sigma$: $49.6^{+ 15.9}_{- 24.8}\pm 6.4$\,$\mu$Jy/beam. All these results are close to those expected from the independently developed SIDES simulations \citep{2017A&A...607A..89B,2022A&A...667A.156B,2023A&A...670A..16G}. This agreement consolidates both our estimates and the model in the flux regime below the detection limit of N2CLS. Since the confusion is caused by a wide range of fluxes, we cannot formally exclude another scenario where the number counts would have a steeper slope than SIDES near the detection limit, but with a higher number of sources with very faint flux densities. However, many such faint galaxies would be required and this would contradict findings that state that low-mass galaxies tend to have lower dust attenuation and, thus, lower FIR emissions \citep[e.g.][]{Heinis2014,Fudamoto2020}.

\begin{acknowledgements}
We acknowledge financial support from the “Programme
National de Cosmologie and Galaxies” (PNCG) funded by CNRS/INSU-IN2P3-INP, CEA and CNES, France, from the European Research Council (ERC) under the European Union’s Horizon 2020 research and innovation programme (project CONCERTO, grant agreement No 788212) and from the Excellence Initiative of Aix-Marseille University-A*Midex, a French “Investissements d’Avenir” programme. This work is based on observations carried out under project numbers
192-16 with the IRAM 30 m telescope. IRAM is supported by INSU/CNRS (France), MPG (Germany) and IGN (Spain). We would like to thank the IRAM staff for their support during the observation campaigns. The NIKA2 dilution cryostat has been designed and built
at the Institut Néel. In particular, we acknowledge the crucial contribution of the Cryogenics Group, and in particular Gregory Garde, Henri Rodenas, Jean-Paul
Leggeri, Philippe Camus. This work has been partially funded by the Foundation Nanoscience Grenoble and the LabEx FOCUS ANR-11-LABX-0013. This work is supported by the French National Research Agency under the contracts “MKIDS”, “NIKA” and ANR-15-CE31-0017 and in the framework of the “Investissements d’avenir” program (ANR-15-IDEX-02). This work has been sup-
ported by the GIS KIDs. This work has benefited from the support of the European Research Council Advanced Grant ORISTARS under the European Union’s
Seventh Framework Programme (Grant Agreement no. 291294). A. R. acknowledges financial support from the Italian Ministry of University and Research –
Project Proposal CIR01\_00010. M.~M.~E.
acknowledges the support of the French Agence Nationale de la Recherche (ANR), under grant ANR-22-CE31-0010. R. A. acknowledges support from the
Programme National Cosmology et Galaxies (PNCG) of CNRS/INSU with INP and IN2P3, co-funded by CEA and CNES. R. A. was supported by the French
government through the France 2030 investment plan managed by the National Research Agency (ANR), as part of the Initiative of Excellence of Universit\'e
Côte d’Azur under reference number ANR-15-IDEX-01.
\end{acknowledgements}

\bibliographystyle{aa}
\bibliography{biblio}

\begin{thebibliography}{50}
\expandafter\ifx\csname natexlab\endcsname\relax\def\natexlab#1{#1}\fi

\bibitem[{{Amblard} {et~al.}(2011){Amblard}, {Cooray}, {Serra}, {Altieri},
  {Arumugam}, {Aussel}, {Blain}, {Bock}, {Boselli}, {Buat},
  {Castro-Rodr{\'{\i}}guez}, {Cava}, {Chanial}, {Chapin}, {Clements}, {Conley},
  {Conversi}, {Dowell}, {Dwek}, {Eales}, {Elbaz}, {Farrah}, {Franceschini},
  {Gear}, {Glenn}, {Griffin}, {Halpern}, {Hatziminaoglou}, {Ibar}, {Isaak},
  {Ivison}, {Khostovan}, {Lagache}, {Levenson}, {Lu}, {Madden}, {Maffei},
  {Mainetti}, {Marchetti}, {Marsden}, {Mitchell-Wynne}, {Nguyen}, {O'Halloran},
  {Oliver}, {Omont}, {Page}, {Panuzzo}, {Papageorgiou}, {Pearson},
  {P{\'e}rez-Fournon}, {Pohlen}, {Rangwala}, {Roseboom}, {Rowan-Robinson},
  {Portal}, {Schulz}, {Scott}, {Seymour}, {Shupe}, {Smith}, {Stevens},
  {Symeonidis}, {Trichas}, {Tugwell}, {Vaccari}, {Valiante}, {Valtchanov},
  {Vieira}, {Vigroux}, {Wang}, {Ward}, {Wright}, {Xu}, \&
  {Zemcov}}]{Amblard2011}
{Amblard}, A., {Cooray}, A., {Serra}, P., {et~al.} 2011, \nat, 470, 510

\bibitem[{{Barger} {et~al.}(1998){Barger}, {Cowie}, {Sanders}, {Fulton},
  {Taniguchi}, {Sato}, {Kawara}, \& {Okuda}}]{Barger1998}
{Barger}, A.~J., {Cowie}, L.~L., {Sanders}, D.~B., {et~al.} 1998, \nat, 394,
  248

\bibitem[{{Berta} {et~al.}(2011){Berta}, {Magnelli}, {Nordon}, {Lutz}, {Wuyts},
  {Altieri}, {Andreani}, {Aussel}, {Casta{\~n}eda}, {Cepa}, {Cimatti}, {Daddi},
  {Elbaz}, {F{\"o}rster Schreiber}, {Genzel}, {Le Floc'h}, {Maiolino},
  {P{\'e}rez-Fournon}, {Poglitsch}, {Popesso}, {Pozzi}, {Riguccini},
  {Rodighiero}, {Sanchez-Portal}, {Sturm}, {Tacconi}, \&
  {Valtchanov}}]{Berta2011}
{Berta}, S., {Magnelli}, B., {Nordon}, R., {et~al.} 2011, \aap, 532, A49

\bibitem[{{B{\'e}thermin} {et~al.}(2024){B{\'e}thermin}, {Bolatto},
  {Boulanger}, {Bradford}, {Burgarella}, {Ciesla}, {Donnellan}, {Hensley},
  {Glenn}, {Lagache}, {Lopez-Rodriguez}, {Oliver}, {Pope}, \&
  {Sauvage}}]{2024A&A...692A..52B}
{B{\'e}thermin}, M., {Bolatto}, A.~D., {Boulanger}, F., {et~al.} 2024, \aap,
  692, A52

\bibitem[{{B{\'e}thermin} {et~al.}(2022){B{\'e}thermin}, {Gkogkou}, {Van
  Cuyck}, {Lagache}, {Beelen}, {Aravena}, {Benoit}, {Bounmy}, {Calvo},
  {Catalano}, {de Batz de Trenquelleon}, {De Breuck}, {Fasano}, {Ferrara},
  {Goupy}, {Hoarau}, {Horellou}, {Hu}, {Julia}, {Knudsen}, {Lambert},
  {Macias-Perez}, {Marpaud}, {Monfardini}, {Pallottini}, {Ponthieu}, {Roehlly},
  {Vallini}, {Walter}, \& {Weiss}}]{2022A&A...667A.156B}
{B{\'e}thermin}, M., {Gkogkou}, A., {Van Cuyck}, M., {et~al.} 2022, \aap, 667,
  A156

\bibitem[{{B{\'e}thermin} {et~al.}(2012){B{\'e}thermin}, {Le Floc'h}, {Ilbert},
  {Conley}, {Lagache}, {Amblard}, {Arumugam}, {Aussel}, {Berta}, {Bock},
  {Boselli}, {Buat}, {Casey}, {Castro-Rodr{\'{\i}}guez}, {Cava}, {Clements},
  {Cooray}, {Dowell}, {Eales}, {Farrah}, {Franceschini}, {Glenn}, {Griffin},
  {Hatziminaoglou}, {Heinis}, {Ibar}, {Ivison}, {Kartaltepe}, {Levenson},
  {Magdis}, {Marchetti}, {Marsden}, {Nguyen}, {O'Halloran}, {Oliver}, {Omont},
  {Page}, {Panuzzo}, {Papageorgiou}, {Pearson}, {P{\'e}rez-Fournon}, {Pohlen},
  {Rigopoulou}, {Roseboom}, {Rowan-Robinson}, {Salvato}, {Schulz}, {Scott},
  {Seymour}, {Shupe}, {Smith}, {Symeonidis}, {Trichas}, {Tugwell}, {Vaccari},
  {Valtchanov}, {Vieira}, {Viero}, {Wang}, {Xu}, \& {Zemcov}}]{Bethermin2012b}
{B{\'e}thermin}, M., {Le Floc'h}, E., {Ilbert}, O., {et~al.} 2012, \aap, 542,
  A58

\bibitem[{{B{\'e}thermin} {et~al.}(2017){B{\'e}thermin}, {Wu}, {Lagache},
  {Davidzon}, {Ponthieu}, {Cousin}, {Wang}, {Dor{\'e}}, {Daddi}, \&
  {Lapi}}]{2017A&A...607A..89B}
{B{\'e}thermin}, M., {Wu}, H.-Y., {Lagache}, G., {et~al.} 2017, \aap, 607, A89

\bibitem[{{Bethermin} {et~al.}(2017){Bethermin}, {Wu}, {Lagache}, {Dor{\'e}},
  {Wang}, \& {Cousin}}]{Bethermin2017}
{Bethermin}, M., {Wu}, H.-Y., {Lagache}, G., {et~al.} 2017, \aap, to be sub.

\bibitem[{{Bing} {et~al.}(2023){Bing}, {B{\'e}thermin}, {Lagache}, {Adam},
  {Ade}, {Ajeddig}, {Andr{\'e}}, {Artis}, {Aussel}, {Beelen}, {Beno{\^\i}t},
  {Berta}, {Billot}, {Bourrion}, {Calvo}, {Catalano}, {De Petris},
  {D{\'e}sert}, {Doyle}, {Driessen}, {Elbaz}, {Gkogkou}, {Gomez}, {Goupy},
  {Hanser}, {K{\'e}ruzor{\'e}}, {Kramer}, {Ladjelate}, {Liu}, {Leclercq},
  {Lestrade}, {Lustig}, {Mac{\'\i}as-P{\'e}rez}, {Maury}, {Mauskopf}, {Mayet},
  {Monfardini}, {Mu{\~n}oz-Echeverr{\'\i}a}, {Perotto}, {Pisano}, {Ponthieu},
  {Rev{\'e}ret}, {Rigby}, {Ritacco}, {Romero}, {Roussel}, {Ruppin}, {Schuster},
  {Sievers}, {Tucker}, \& {Zylka}}]{bing}
{Bing}, L., {B{\'e}thermin}, M., {Lagache}, G., {et~al.} 2023, \aap, 677, A66

\bibitem[{{Blain} {et~al.}(1998){Blain}, {Ivison}, \& {Smail}}]{Blain1998}
{Blain}, A.~W., {Ivison}, R.~J., \& {Smail}, I. 1998, \mnras, 296, L29

\bibitem[{{Calvo} {et~al.}(2013){Calvo}, {Roesch}, {D{\'e}sert}, {Monfardini},
  {Benoit}, {Mauskopf}, {Ade}, {Boudou}, {Bourrion}, {Camus}, {Cruciani},
  {Doyle}, {Hoffmann}, {Leclercq}, {Macias-Perez}, {Ponthieu}, {Schuster},
  {Tucker}, \& {Vescovi}}]{2013A&A...551L..12C}
{Calvo}, M., {Roesch}, M., {D{\'e}sert}, F.~X., {et~al.} 2013, \aap, 551, L12

\bibitem[{{Catalano} {et~al.}(2014){Catalano}, {Calvo}, {Ponthieu}, {Adam},
  {Adane}, {Ade}, {Andr{\'e}}, {Beelen}, {Belier}, {Beno{\^\i}t}, {Bideaud},
  {Billot}, {Boudou}, {Bourrion}, {Coiffard}, {Comis}, {D'Addabbo},
  {D{\'e}sert}, {Doyle}, {Goupy}, {Kramer}, {Leclercq},
  {Mac{\'\i}as-P{\'e}rez}, {Martino}, {Mauskopf}, {Mayet}, {Monfardini},
  {Pajot}, {Pascale}, {Perotto}, {Rev{\'e}ret}, {Rodriguez}, {Savini},
  {Schuster}, {Sievers}, {Tucker}, \& {Zylka}}]{Catalano2014}
{Catalano}, A., {Calvo}, M., {Ponthieu}, N., {et~al.} 2014, \aap, 569, A9

\bibitem[{{Condon}(1974)}]{Condon1974}
{Condon}, J.~J. 1974, \apj, 188, 279

\bibitem[{{Dole} {et~al.}(2003{\natexlab{a}}){Dole}, {Lagache}, \&
  {Puget}}]{Dole2003}
{Dole}, H., {Lagache}, G., \& {Puget}, J. 2003{\natexlab{a}}, \apj, 585, 617

\bibitem[{{Dole} {et~al.}(2006){Dole}, {Lagache}, {Puget}, {Caputi},
  {Fern{\'a}ndez-Conde}, {Le Floc'h}, {Papovich}, {P{\'e}rez-Gonz{\'a}lez},
  {Rieke}, \& {Blaylock}}]{Dole2006}
{Dole}, H., {Lagache}, G., {Puget}, J., {et~al.} 2006, \aap, 451, 417

\bibitem[{{Dole} {et~al.}(2003{\natexlab{b}}){Dole}, {Lagache}, \&
  {Puget}}]{Dole+03}
{Dole}, H., {Lagache}, G., \& {Puget}, J.~L. 2003{\natexlab{b}}, \apj, 585, 617

\bibitem[{{Dole} {et~al.}(2004){Dole}, {Rieke}, {Lagache}, {Puget},
  {Alonso-Herrero}, {Bai}, {Blaylock}, {Egami}, {Engelbracht}, {Gordon},
  {Hines}, {Kelly}, {Le Floc'h}, {Misselt}, {Morrison}, {Muzerolle},
  {Papovich}, {P{\'e}rez-Gonz{\'a}lez}, {Rieke}, {Rigby}, {Neugebauer},
  {Stansberry}, {Su}, {Young}, {Beichman}, \& {Richards}}]{Dole+04}
{Dole}, H., {Rieke}, G.~H., {Lagache}, G., {et~al.} 2004, \apjs, 154, 93

\bibitem[{{Eales} {et~al.}(1999){Eales}, {Lilly}, {Gear}, {Dunne}, {Bond},
  {Hammer}, {Le F{\`e}vre}, \& {Crampton}}]{Eales1999}
{Eales}, S., {Lilly}, S., {Gear}, W., {et~al.} 1999, \apj, 515, 518

\bibitem[{{Frayer} {et~al.}(2006){Frayer}, {Huynh}, {Chary}, {Dickinson},
  {Elbaz}, {Fadda}, {Surace}, {Teplitz}, {Yan}, \& {Mobasher}}]{Frayer2006b}
{Frayer}, D.~T., {Huynh}, M.~T., {Chary}, R., {et~al.} 2006, \apjl, 647, L9

\bibitem[{{Fudamoto} {et~al.}(2020){Fudamoto}, {Oesch}, {Faisst},
  {B{\'e}thermin}, {Ginolfi}, {Khusanova}, {Loiacono}, {Le F{\`e}vre}, {Capak},
  {Schaerer}, {Silverman}, {Cassata}, {Yan}, {Amorin}, {Bardelli}, {Boquien},
  {Cimatti}, {Dessauges-Zavadsky}, {Fujimoto}, {Gruppioni}, {Hathi}, {Ibar},
  {Jones}, {Koekemoer}, {Lagache}, {Lemaux}, {Maiolino}, {Narayanan}, {Pozzi},
  {Riechers}, {Rodighiero}, {Talia}, {Toft}, {Vallini}, {Vergani}, {Zamorani},
  \& {Zucca}}]{Fudamoto2020}
{Fudamoto}, Y., {Oesch}, P.~A., {Faisst}, A., {et~al.} 2020, \aap, 643, A4

\bibitem[{{Gautier} {et~al.}(1992){Gautier}, {Boulanger}, {Perault}, \&
  {Puget}}]{Gautier+92}
{Gautier}, T.~N., I., {Boulanger}, F., {Perault}, M., \& {Puget}, J.~L. 1992,
  \aj, 103, 1313

\bibitem[{{Geach} {et~al.}(2017){Geach}, {Dunlop}, {Halpern}, {Smail}, {van der
  Werf}, {Alexander}, {Almaini}, {Aretxaga}, {Arumugam}, {Asboth}, {Banerji},
  {Beanlands}, {Best}, {Blain}, {Birkinshaw}, {Chapin}, {Chapman}, {Chen},
  {Chrysostomou}, {Clarke}, {Clements}, {Conselice}, {Coppin}, {Cowley},
  {Danielson}, {Eales}, {Edge}, {Farrah}, {Gibb}, {Harrison}, {Hine}, {Hughes},
  {Ivison}, {Jarvis}, {Jenness}, {Jones}, {Karim}, {Koprowski}, {Knudsen},
  {Lacey}, {Mackenzie}, {Marsden}, {McAlpine}, {McMahon}, {Meijerink},
  {Micha{\l}owski}, {Oliver}, {Page}, {Peacock}, {Rigopoulou}, {Robson},
  {Roseboom}, {Rotermund}, {Scott}, {Serjeant}, {Simpson}, {Simpson}, {Smith},
  {Spaans}, {Stanley}, {Stevens}, {Swinbank}, {Targett}, {Thomson}, {Valiante},
  {Wake}, {Webb}, {Willott}, {Zavala}, \& {Zemcov}}]{Geach2017}
{Geach}, J.~E., {Dunlop}, J.~S., {Halpern}, M., {et~al.} 2017, \mnras, 465,
  1789

\bibitem[{{Gkogkou} {et~al.}(2023){Gkogkou}, {B{\'e}thermin}, {Lagache}, {Van
  Cuyck}, {Jullo}, {Aravena}, {Beelen}, {Benoit}, {Bounmy}, {Calvo},
  {Catalano}, {Cora}, {Croton}, {de la Torre}, {Fasano}, {Ferrara}, {Goupy},
  {Hoarau}, {Hu}, {Ishiyama}, {Knudsen}, {Lambert}, {Mac{\'\i}as-P{\'e}rez},
  {Marpaud}, {Mellema}, {Monfardini}, {Pallottini}, {Ponthieu}, {Prada},
  {Roehlly}, {Vallini}, \& {Walter}}]{2023A&A...670A..16G}
{Gkogkou}, A., {B{\'e}thermin}, M., {Lagache}, G., {et~al.} 2023, \aap, 670,
  A16

\bibitem[{{Glenn} {et~al.}(2010){Glenn}, {Conley}, {B{\'e}thermin}, {Altieri},
  {Amblard}, {Arumugam}, {Aussel}, {Babbedge}, {Blain}, {Bock}, {Boselli},
  {Buat}, {Castro-Rodr{\'{\i}}guez}, {Cava}, {Chanial}, {Clements}, {Conversi},
  {Cooray}, {Dowell}, {Dwek}, {Eales}, {Elbaz}, {Ellsworth-Bowers}, {Fox},
  {Franceschini}, {Gear}, {Griffin}, {Halpern}, {Hatziminaoglou}, {Ibar},
  {Isaak}, {Ivison}, {Lagache}, {Laurent}, {Levenson}, {Lu}, {Madden},
  {Maffei}, {Mainetti}, {Marchetti}, {Marsden}, {Nguyen}, {O'Halloran},
  {Oliver}, {Omont}, {Page}, {Panuzzo}, {Papageorgiou}, {Pearson},
  {P{\'e}rez-Fournon}, {Pohlen}, {Rigopoulou}, {Rizzo}, {Roseboom},
  {Rowan-Robinson}, {Portal}, {Schulz}, {Scott}, {Seymour}, {Shupe}, {Smith},
  {Stevens}, {Symeonidis}, {Trichas}, {Tugwell}, {Vaccari}, {Valtchanov},
  {Vieira}, {Vigroux}, {Wang}, {Ward}, {Wright}, {Xu}, \& {Zemcov}}]{Glenn2010}
{Glenn}, J., {Conley}, A., {B{\'e}thermin}, M., {et~al.} 2010, \mnras, 409, 109

\bibitem[{{Grabovskij} {et~al.}(2008){Grabovskij}, {Swenson}, {Buisson},
  {Hoffmann}, {Monfardini}, \& {Vill{\'e}gier}}]{2008ApPhL..93m4102G}
{Grabovskij}, G.~J., {Swenson}, L.~J., {Buisson}, O., {et~al.} 2008, Applied
  Physics Letters, 93, 134102

\bibitem[{{Hacking} \& {Soifer}(1991)}]{Hacking1991}
{Hacking}, P.~B. \& {Soifer}, B.~T. 1991, \apjl, 367, L49

\bibitem[{{Heinis} {et~al.}(2014){Heinis}, {Buat}, {B{\'e}thermin}, {Bock},
  {Burgarella}, {Conley}, {Cooray}, {Farrah}, {Ilbert}, {Magdis}, {Marsden},
  {Oliver}, {Rigopoulou}, {Roehlly}, {Schulz}, {Symeonidis}, {Viero}, {Xu}, \&
  {Zemcov}}]{Heinis2014}
{Heinis}, S., {Buat}, V., {B{\'e}thermin}, M., {et~al.} 2014, \mnras, 437, 1268

\bibitem[{{Helou} \& {Beichman}(1990)}]{Helou1990}
{Helou}, G. \& {Beichman}, C.~A. 1990, in Liege International Astrophysical
  Colloquia, Vol.~29, Liege International Astrophysical Colloquia, ed.
  B.~{Kaldeich}, 117--123

\bibitem[{{Hughes} {et~al.}(1998){Hughes}, {Serjeant}, {Dunlop},
  {Rowan-Robinson}, {Blain}, {Mann}, {Ivison}, {Peacock}, {Efstathiou}, {Gear},
  {Oliver}, {Lawrence}, {Longair}, {Goldschmidt}, \& {Jenness}}]{Hughes1998}
{Hughes}, D.~H., {Serjeant}, S., {Dunlop}, J., {et~al.} 1998, \nat, 394, 241

\bibitem[{{Kiss} {et~al.}(2001){Kiss}, {{\'A}brah{\'a}m}, {Klaas}, {Juvela}, \&
  {Lemke}}]{Kiss+01}
{Kiss}, C., {{\'A}brah{\'a}m}, P., {Klaas}, U., {Juvela}, M., \& {Lemke}, D.
  2001, \aap, 379, 1161

\bibitem[{{Kiss} {et~al.}(2005){Kiss}, {Klaas}, \& {Lemke}}]{Kiss+05}
{Kiss}, C., {Klaas}, U., \& {Lemke}, D. 2005, \aap, 430, 343

\bibitem[{{Lagache} {et~al.}(2003){Lagache}, {Dole}, \& {Puget}}]{Lagache+03}
{Lagache}, G., {Dole}, H., \& {Puget}, J.~L. 2003, \mnras, 338, 555

\bibitem[{{Lagache} {et~al.}(2000){Lagache}, {Haffner}, {Reynolds}, \&
  {Tufte}}]{Lagache2000}
{Lagache}, G., {Haffner}, L.~M., {Reynolds}, R.~J., \& {Tufte}, S.~L. 2000,
  \aap, 354, 247

\bibitem[{{Ledden} {et~al.}(1980){Ledden}, {Broderick}, {Condon}, \&
  {Brown}}]{Ledden1980}
{Ledden}, J.~E., {Broderick}, J.~J., {Condon}, J.~J., \& {Brown}, R.~L. 1980,
  \aj, 85, 780

\bibitem[{{Monfardini} {et~al.}(2014){Monfardini}, {Adam}, {Adane}, {Ade},
  {Andr{\'e}}, {Beelen}, {Belier}, {Benoit}, {Bideaud}, {Billot}, {Bourrion},
  {Calvo}, {Catalano}, {Coiffard}, {Comis}, {D'Addabbo}, {D{\'e}sert}, {Doyle},
  {Goupy}, {Kramer}, {Leclercq}, {Macias-Perez}, {Martino}, {Mauskopf},
  {Mayet}, {Pajot}, {Pascale}, {Ponthieu}, {Rev{\'e}ret}, {Rodriguez},
  {Savini}, {Schuster}, {Sievers}, {Tucker}, \& {Zylka}}]{Monfardini2014}
{Monfardini}, A., {Adam}, R., {Adane}, A., {et~al.} 2014, Journal of Low
  Temperature Physics, 176, 787

\bibitem[{{Monfardini} {et~al.}(2010){Monfardini}, {Swenson}, {Bideaud},
  {D{\'e}sert}, {Yates}, {Benoit}, {Baryshev}, {Baselmans}, {Doyle}, {Klein},
  {Roesch}, {Tucker}, {Ade}, {Calvo}, {Camus}, {Giordano}, {Guesten},
  {Hoffmann}, {Leclercq}, {Mauskopf}, \& {Schuster}}]{Monfardini2010}
{Monfardini}, A., {Swenson}, L.~J., {Bideaud}, A., {et~al.} 2010, \aap, 521,
  A29

\bibitem[{{Negrello} {et~al.}(2004){Negrello}, {Magliocchetti}, {Moscardini},
  {De Zotti}, {Granato}, \& {Silva}}]{Negrello+04}
{Negrello}, M., {Magliocchetti}, M., {Moscardini}, L., {et~al.} 2004, \mnras,
  352, 493

\bibitem[{{Nguyen} {et~al.}(2010{\natexlab{a}}){Nguyen}, {Schulz}, {Levenson},
  {Amblard}, {Arumugam}, {Aussel}, {Babbedge}, {Blain}, {Bock}, {Boselli},
  {Buat}, {Castro-Rodriguez}, {Cava}, {Chanial}, {Chapin}, {Clements},
  {Conley}, {Conversi}, {Cooray}, {Dowell}, {Dwek}, {Eales}, {Elbaz}, {Fox},
  {Franceschini}, {Gear}, {Glenn}, {Griffin}, {Halpern}, {Hatziminaoglou},
  {Ibar}, {Isaak}, {Ivison}, {Lagache}, {Lu}, {Madden}, {Maffei}, {Mainetti},
  {Marchetti}, {Marsden}, {Marshall}, {O'Halloran}, {Oliver}, {Omont}, {Page},
  {Panuzzo}, {Papageorgiou}, {Pearson}, {Perez Fournon}, {Pohlen}, {Rangwala},
  {Rigopoulou}, {Rizzo}, {Roseboom}, {Rowan-Robinson}, {Scott}, {Seymour},
  {Shupe}, {Smith}, {Stevens}, {Symeonidis}, {Trichas}, {Tugwell}, {Vaccari},
  {Valtchanov}, {Vigroux}, {Wang}, {Ward}, {Wiebe}, {Wright}, {Xu}, \&
  {Zemcov}}]{Nguyen2010}
{Nguyen}, H.~T., {Schulz}, B., {Levenson}, L., {et~al.} 2010{\natexlab{a}},
  \aap, 518, L5

\bibitem[{{Nguyen} {et~al.}(2010{\natexlab{b}}){Nguyen}, {Schulz}, {Levenson},
  {Amblard}, {Arumugam}, {Aussel}, {Babbedge}, {Blain}, {Bock}, {Boselli},
  {Buat}, {Castro-Rodriguez}, {Cava}, {Chanial}, {Chapin}, {Clements},
  {Conley}, {Conversi}, {Cooray}, {Dowell}, {Dwek}, {Eales}, {Elbaz}, {Fox},
  {Franceschini}, {Gear}, {Glenn}, {Griffin}, {Halpern}, {Hatziminaoglou},
  {Ibar}, {Isaak}, {Ivison}, {Lagache}, {Lu}, {Madden}, {Maffei}, {Mainetti},
  {Marchetti}, {Marsden}, {Marshall}, {O'Halloran}, {Oliver}, {Omont}, {Page},
  {Panuzzo}, {Papageorgiou}, {Pearson}, {Perez Fournon}, {Pohlen}, {Rangwala},
  {Rigopoulou}, {Rizzo}, {Roseboom}, {Rowan-Robinson}, {Scott}, {Seymour},
  {Shupe}, {Smith}, {Stevens}, {Symeonidis}, {Trichas}, {Tugwell}, {Vaccari},
  {Valtchanov}, {Vigroux}, {Wang}, {Ward}, {Wiebe}, {Wright}, {Xu}, \&
  {Zemcov}}]{2010A&A...518L...5N}
{Nguyen}, H.~T., {Schulz}, B., {Levenson}, L., {et~al.} 2010{\natexlab{b}},
  \aap, 518, L5

\bibitem[{{Oliver} {et~al.}(2010){Oliver}, {Wang}, {Smith}, {Altieri},
  {Amblard}, {Arumugam}, {Auld}, {Aussel}, {Babbedge}, {Blain}, {Bock},
  {Boselli}, {Buat}, {Burgarella}, {Castro-Rodr{\'{\i}}guez}, {Cava},
  {Chanial}, {Clements}, {Conley}, {Conversi}, {Cooray}, {Dowell}, {Dwek},
  {Eales}, {Elbaz}, {Fox}, {Franceschini}, {Gear}, {Glenn}, {Griffin},
  {Halpern}, {Hatziminaoglou}, {Ibar}, {Isaak}, {Ivison}, {Lagache},
  {Levenson}, {Lu}, {Madden}, {Maffei}, {Mainetti}, {Marchetti},
  {Mitchell-Wynne}, {Mortier}, {Nguyen}, {O'Halloran}, {Omont}, {Page},
  {Panuzzo}, {Papageorgiou}, {Pearson}, {P{\'e}rez-Fournon}, {Pohlen},
  {Rawlings}, {Raymond}, {Rigopoulou}, {Rizzo}, {Roseboom}, {Rowan-Robinson},
  {S{\'a}nchez Portal}, {Savage}, {Schulz}, {Scott}, {Seymour}, {Shupe},
  {Stevens}, {Symeonidis}, {Trichas}, {Tugwell}, {Vaccari}, {Valiante},
  {Valtchanov}, {Vieira}, {Vigroux}, {Ward}, {Wright}, {Xu}, \&
  {Zemcov}}]{Oliver2010}
{Oliver}, S.~J., {Wang}, L., {Smith}, A.~J., {et~al.} 2010, \aap, 518, L21+

\bibitem[{{Papovich} {et~al.}(2004){Papovich}, {Dole}, {Egami}, {Le Floc'h},
  {P{\'e}rez-Gonz{\'a}lez}, {Alonso-Herrero}, {Bai}, {Beichman}, {Blaylock},
  {Engelbracht}, {Gordon}, {Hines}, {Misselt}, {Morrison}, {Mould},
  {Muzerolle}, {Neugebauer}, {Richards}, {Rieke}, {Rieke}, {Rigby}, {Su}, \&
  {Young}}]{Papovich2004}
{Papovich}, C., {Dole}, H., {Egami}, E., {et~al.} 2004, \apjs, 154, 70

\bibitem[{{Perotto} {et~al.}(2020){Perotto}, {Ponthieu},
  {Mac{\'\i}as-P{\'e}rez}, {Adam}, {Ade}, {Andr{\'e}}, {Andrianasolo},
  {Aussel}, {Beelen}, {Beno{\^\i}t}, {Berta}, {Bideaud}, {Bourrion}, {Calvo},
  {Catalano}, {Comis}, {De Petris}, {D{\'e}sert}, {Doyle}, {Driessen},
  {Garc{\'\i}a}, {Gomez}, {Goupy}, {John}, {K{\'e}ruzor{\'e}}, {Kramer},
  {Ladjelate}, {Lagache}, {Leclercq}, {Lestrade}, {Maury}, {Mauskopf}, {Mayet},
  {Monfardini}, {Navarro}, {Pe{\~n}alver}, {Pierfederici}, {Pisano},
  {Rev{\'e}ret}, {Ritacco}, {Romero}, {Roussel}, {Ruppin}, {Schuster}, {Shu},
  {Sievers}, {Tucker}, \& {Zylka}}]{perotto20}
{Perotto}, L., {Ponthieu}, N., {Mac{\'\i}as-P{\'e}rez}, J.~F., {et~al.} 2020,
  \aap, 637, A71

\bibitem[{{Pilbratt} {et~al.}(2010){Pilbratt}, {Riedinger}, {Passvogel},
  {Crone}, {Doyle}, {Gageur}, {Heras}, {Jewell}, {Metcalfe}, {Ott}, \&
  {Schmidt}}]{Pilbratt2010}
{Pilbratt}, G.~L., {Riedinger}, J.~R., {Passvogel}, T., {et~al.} 2010, \aap,
  518, L1+

\bibitem[{{Planck Collaboration} {et~al.}(2014){Planck Collaboration}, {Ade},
  {Aghanim}, {Armitage-Caplan}, {Arnaud}, {Ashdown}, {Atrio-Barandela},
  {Aumont}, {Baccigalupi}, {Banday}, \& et~al.}]{Planck_CIB2013}
{Planck Collaboration}, {Ade}, P.~A.~R., {Aghanim}, N., {et~al.} 2014, \aap,
  571, A30

\bibitem[{{Planck Collaboration} {et~al.}(2011){Planck Collaboration}, {Ade},
  {Aghanim}, {Arnaud}, {Ashdown}, {Aumont}, {Baccigalupi}, {Balbi}, {Banday},
  {Barreiro}, \& et~al.}]{Planck_CIB2011}
{Planck Collaboration}, {Ade}, P.~A.~R., {Aghanim}, N., {et~al.} 2011, \aap,
  536, A18

\bibitem[{{Takeuchi} {et~al.}(2001){Takeuchi}, {Kawabe}, {Kohno}, {Nakanishi},
  {Ishii}, {Hirashita}, \& {Yoshikawa}}]{Takeuchi2001}
{Takeuchi}, T.~T., {Kawabe}, R., {Kohno}, K., {et~al.} 2001, \pasp, 113, 586

\bibitem[{{Tristram} {et~al.}(2005){Tristram}, {Mac{\'\i}as-P{\'e}rez},
  {Renault}, \& {Santos}}]{2005MNRAS.358..833T}
{Tristram}, M., {Mac{\'\i}as-P{\'e}rez}, J.~F., {Renault}, C., \& {Santos}, D.
  2005, \mnras, 358, 833

\bibitem[{{Viero} {et~al.}(2013{\natexlab{a}}){Viero}, {Moncelsi}, {Quadri},
  {Arumugam}, {Assef}, {B{\'e}thermin}, {Bock}, {Bridge}, {Casey}, {Conley},
  {Cooray}, {Farrah}, {Glenn}, {Heinis}, {Ibar}, {Ikarashi}, {Ivison}, {Kohno},
  {Marsden}, {Oliver}, {Roseboom}, {Schulz}, {Scott}, {Serra}, {Vaccari},
  {Vieira}, {Wang}, {Wardlow}, {Wilson}, {Yun}, \& {Zemcov}}]{Viero2013b}
{Viero}, M.~P., {Moncelsi}, L., {Quadri}, R.~F., {et~al.} 2013{\natexlab{a}},
  \apj, 779, 32

\bibitem[{{Viero} {et~al.}(2013{\natexlab{b}}){Viero}, {Wang}, {Zemcov},
  {Addison}, {Amblard}, {Arumugam}, {Aussel}, {B{\'e}thermin}, {Bock},
  {Boselli}, {Buat}, {Burgarella}, {Casey}, {Clements}, {Conley}, {Conversi},
  {Cooray}, {De Zotti}, {Dowell}, {Farrah}, {Franceschini}, {Glenn}, {Griffin},
  {Hatziminaoglou}, {Heinis}, {Ibar}, {Ivison}, {Lagache}, {Levenson},
  {Marchetti}, {Marsden}, {Nguyen}, {O'Halloran}, {Oliver}, {Omont}, {Page},
  {Papageorgiou}, {Pearson}, {P{\'e}rez-Fournon}, {Pohlen}, {Rigopoulou},
  {Roseboom}, {Rowan-Robinson}, {Schulz}, {Scott}, {Seymour}, {Shupe}, {Smith},
  {Symeonidis}, {Vaccari}, {Valtchanov}, {Vieira}, {Wardlow}, \&
  {Xu}}]{Viero2013}
{Viero}, M.~P., {Wang}, L., {Zemcov}, M., {et~al.} 2013{\natexlab{b}}, \apj,
  772, 77

\bibitem[{{Werner} {et~al.}(2004){Werner}, {Roellig}, {Low}, {Rieke}, {Rieke},
  {Hoffmann}, {Young}, {Houck}, {Brandl}, {Fazio}, {Hora}, {Gehrz}, {Helou},
  {Soifer}, {Stauffer}, {Keene}, {Eisenhardt}, {Gallagher}, {Gautier}, {Irace},
  {Lawrence}, {Simmons}, {Van Cleve}, {Jura}, {Wright}, \&
  {Cruikshank}}]{Werner2004}
{Werner}, M.~W., {Roellig}, T.~L., {Low}, F.~J., {et~al.} 2004, \apjs, 154, 1

\end{thebibliography}

\clearpage
\appendix

\section{Estimation of the confusion via the \xrms}
\label{app:equations}

To motivate the definition of our estimator, we start with the simple case where the signal projected onto maps is unaffected by the data processing and the noise in the map pixels is Gaussian, white, and independent per run. We then return to these hypotheses.

We consider two different runs of observations. The combination of all scans for the first gives the map $A$, the combination of all scans for the second gives the map $B$. In the following, capital letters denote map raw values and $p$ and $q$ denote pixel indices. Let $S_p$ be the sky signal at pixel $p$, so for map $A$ we have $A_p = S_p + a_p$, where $a_p$ is the noise of map $A$. We assume that masking bright sources leaves us with $N$ unmasked pixels. Finally, we use an over bar to denote the spatial average of the sky signal over all these pixels:

\begin{equation}
\bar{S} = \frac{1}{N}\sum_{p=1}^N S_p, \nonumber
\end{equation}
With such conventions, the confusion at 1 or 2\,mm is expressed as
\begin{equation}
\sigma_c^2 = \frac{1}{N}\sum_p (S_p-\bar{S})^2\,.
\end{equation}

We generalise this definition to the  cross-band confusion when $A$ is a 1\,mm map and $B$ a 2\,mm map.
\begin{equation}
\sigma_{c_{1\times 2}}^2 = \frac{1}{N}\sum_p (S^{1\rm{mm}}_p-\bar{S}^{1\rm{mm}})(S^{2\rm{mm}}_p-\bar{S}^{2\rm{mm}}),
\end{equation}

Under the hypotheses of pure signal and ideal noise, we define the ideal cross-variance estimator as
\begin{equation}
\tilde{\rho}_{AB} = \frac{1}{N}\sum_p (A_p-\bar{A})(B_p-\bar{B}),
\label{eq:rho_ab}
\end{equation}

We show that its statistical expectancy is indeed $\sigma_c^2$ via

\begin{eqnarray}
\langle \tilde{\rho}_{AB}\rangle &= & \frac{1}{N}\sum_p  \langle (A_p-\bar{A})(B_p-\bar{B})\rangle \nonumber \\
&=& \frac{1}{N}\sum_p \langle (S^A_p-\bar{S}^A+a_p)(S^B_p-\bar{S}^B+b_p)\rangle \nonumber\\
&=& \frac{1}{N}\sum_p (S^A_p-\bar{S}^A)(S^B_p-\bar{S}^B) \label{eq:exp_rho} \nonumber \\
&=& \sigma_c^2 \; \rm{or}\; \sigma_{c_{1\times 2}}^2
\label{eq:rho_expect}
.\end{eqnarray}

Next, we can compute the covariance of this estimator. By definition, taking four maps, $A$, $B$, $C,$ and $D$, the covariance matrix of $\tilde{\rho}$ is expressed as

\begin{eqnarray}
\tilde{X}_{AB,CD} & = & \biggl<(\tilde{\rho}_{AB}-\langle\tilde{\rho}_{AB}\rangle)(\tilde{\rho}_{CD}-\langle\tilde{\rho}_{CD}\rangle)\biggr> \nonumber \\
&=&
 \biggl< \tilde{\rho}_{AB}\tilde{\rho}_{CD} - \tilde{\rho}_{AB}\langle\tilde{\rho}_{CD}\rangle - \langle\tilde{\rho}_{AB}\rangle\tilde{\rho}_{CD} + \langle\tilde{\rho}_{AB}\rangle \langle\tilde{\rho}_{CD}\rangle\biggr>.\nonumber\\
  \label{eq:covar_abcd_def}
 \end{eqnarray}
 
We  define $T_p = S_p-\bar{S}$ for convenience and use Kronecker's symbol $\delta, $  which is equal to 1 when its indices are equal, zero otherwise. The first of the four terms in Eq.~(\ref{eq:covar_abcd_def}) is expressed as 
 \begin{eqnarray}
 \langle \tilde{\rho}_{AB}\tilde{\rho}_{CD}\rangle  &=& 
\frac{1}{N^2} \biggl<  \sum_{p,q} (A_p-\bar{A})(B_p-\bar{B}) (C_q-\bar{C})(D_q-\bar{D}) \biggr> \nonumber \\
&=&
\frac{1}{N^2} \biggl<  \sum_{p,q} (S^A_p-\bar{S}^A+a_p)(S^B_p-\bar{S}^B+b_p)\nonumber\\
&& (S^C_q-\bar{S}^C+c_q)(S^D_q-\bar{S}^D+d_q) \biggr> \nonumber \\
&=&
\frac{1}{N^2} \biggl<  \sum_{p,q} (T^A_p+a_p)(T^B_p+b_p)(T^C_q+c_q)(T^D_q+d_q) \biggr> \nonumber\\
&=&
\frac{1}{N^2} \biggl<  \sum_{p,q} \left(T^A_pT^B_p + T^A_p b_p + T^B_p a_p +a_pb_p\right)\nonumber\\
&&\left(T^C_q T^D_q + T^C_q d_q + T^D_q c_q + c_qd_q\right)\biggr> \nonumber \\
&=&
\frac{1}{N^2} \sum_{p,q} \delta_{pq}\left[ T^A_pT^B_pT^C_q T^D_q + \right. \nonumber \\
&& (T^A_pT^C_q\delta_{BD} + T^A_pT^D_q\delta_{BC})\sigma^2_{B_p} + \nonumber\\
&& (T^B_pT^C_q\delta_{AD}+T^B_pT^D_q\delta_{AC})\sigma^2_{A_p}+ \nonumber\\
&&  \left. (\delta_{AD}\delta_{BC} + \delta_{AC}\delta_{BD})\sigma^2_{A_p}\sigma^2_{B_p}\right]\,. \label{eq:rho_abcd}
\end{eqnarray}

In the previous equation, $\sigma^2_{A_p}$ for instance, denotes the variance of the noise $a_p$ in pixel $p$ of map $A$. It is derived at the map making stage when all scans are co-added (Sect.~\ref{se:map_making}) and will be responsible for the statistical uncertainty on the determination of the confusion.

Assuming for now that $A$, $B$, $C,$ and $D$ are all maps of the same band, then Eq.~(\ref{eq:rho_abcd}) becomes:

\begin{eqnarray}
\langle \tilde{\rho}_{AB}\tilde{\rho}_{CD}\rangle &=& \frac{1}{N^2} \sum_{pq} \delta_{pq}\left[ T_p^2T_q^2 \right. \nonumber\\
&&+ T_p^2\left[(\delta_{BD} + \delta_{BC})\sigma^2_{B_p} + (\delta_{AD}+\delta_{AC})\sigma^2_{A_p}\right]\nonumber\\
&&+(\delta_{AD}\delta_{BC} + \delta_{AC}\delta_{BD})\sigma^2_{A_p}\sigma^2_{B_p}
\end{eqnarray}

The other terms involved in Eq.~(\ref{eq:covar_abcd_def}) are
 \begin{eqnarray}
\biggl<\tilde{\rho}_{AB}\langle\tilde{\rho}_{CD}\rangle\biggr>&=&\langle \tilde{\rho}_{AB}\rangle \frac{1}{N}\sum_q T^C_qT^D_q \nonumber\\
&=&
\frac{1}{N}\sum_p\biggl<T^A_pT^B_p + T^A_pb_p+T^B_pa_p+ a_pb_p\biggr>\nonumber\\
&&\times \frac{1}{N}\sum_q T^C_qT^D_q \nonumber\\
&=&
\frac{1}{N^2}\sum_{pq}T^A_pT^B_pT^C_qT^D_q\,.
\end{eqnarray}

The same goes for the last two terms:
 \begin{equation}
\biggl< \langle\tilde{\rho}_{AB}\rangle\tilde{\rho}_{CD}\biggr> = \frac{1}{N^2}\sum_{pq}T^A_pT^B_pT^C_qT^D_q = \langle\tilde{\rho}_{AB}\rangle \langle \tilde{\rho}_{CD}\rangle\,, \nonumber
\end{equation}

hence,

\begin{eqnarray}
\tilde{X}^{1\times1 \;{\rm or}\; 2\times2}_{AB,CD} &=&\frac{1}{N^2}\sum_p T_p^2\left[(\delta_{AC}+\delta_{AD})\sigma_{A_p}^2 + (\delta_{BC}+\delta_{BD})\sigma_{B_p}^2\right]\nonumber\\
&&+ (\delta_{AC}\delta_{BD} + \delta_{AD}\delta_{BC})\sigma_{A_p}^2\sigma_{B_p}^2\,.
\label{eq:x_abcd_1}
\end{eqnarray}

In the specific case of the cross-band confusion, because we take maps from different runs and bands, if $A$ and $C$ refer to maps at 1\,mm and $B$ and $D$ to maps at 2\,mm, then eq.~(\ref{eq:x_abcd_1}) simplifies further into

\begin{eqnarray}
\tilde{X}^{1\times 2}_{AB,CD} &=&\frac{1}{N^2}\sum_p T_p^2(1\,{\rm mm}) \delta_{BD}\sigma_{B_p}^2 + T^2_p(2\,{\rm mm})\delta_{AC}\sigma_{A_p}^2 \nonumber\\
&&+ \delta_{AC}\delta_{BD}\sigma_{A_p}^2\sigma_{B_p}^2
\label{eq:x_abcd_12}
,\end{eqnarray}

We note that in each case, the covariance matrix involves terms like $T_p^2$, that is to say the pure underlying signal, that we actually don't know in real data. In practice, we replace this term by the expected confusion squared $\sigma_c^2$. Because it is in any case much smaller than the noise per pixel, the uncertainty on its exact value has a negligible impact on the derivation of the covariance matrix and subsequently that of the final error bar.

In a more realistic case, the data are inevitably affected by the data processing. It is common to loosely refer to the overall effect as "filtering". In this work, where we determine a single scalar term, we model the impact of the filtering by a multiplicative factor $f$ on the confusion, smaller than 1, that needs to be determined. Thus, our \xrms\ estimator and its covariance become:

\begin{eqnarray}
    \rho_{AB} &=& \frac{1}{f^2}\,\tilde{\rho}_{AB} \label{eq:f_and_rho} \\
    X_{AB,CD} &=& \frac{1}{f^4}\,\tilde{X}_{AB,CD}
\end{eqnarray}

\begin{figure}
\centering
\includegraphics[width=9cm]{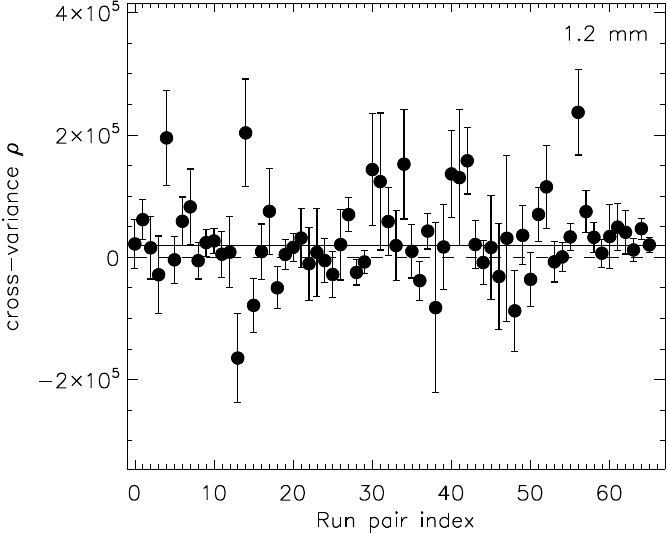}
\caption{Estimates of the \xrms\ at 1.2\,mm for all pairs of independent observation runs in $(\mu$Jy/beam)$^2$. Error bars are the square root of the diagonal elements of the covariance matrix of $\rho_i$ shown on Fig.~\ref{fig:rho_covar_matrix}. The dashed line marks zero and the solid line is the combined \xrms\ (see Eq.~(\ref{eq:opt_rho})).}
\label{fig:all_rho}
\end{figure}

\begin{figure}
\centering
\includegraphics[width=9cm]{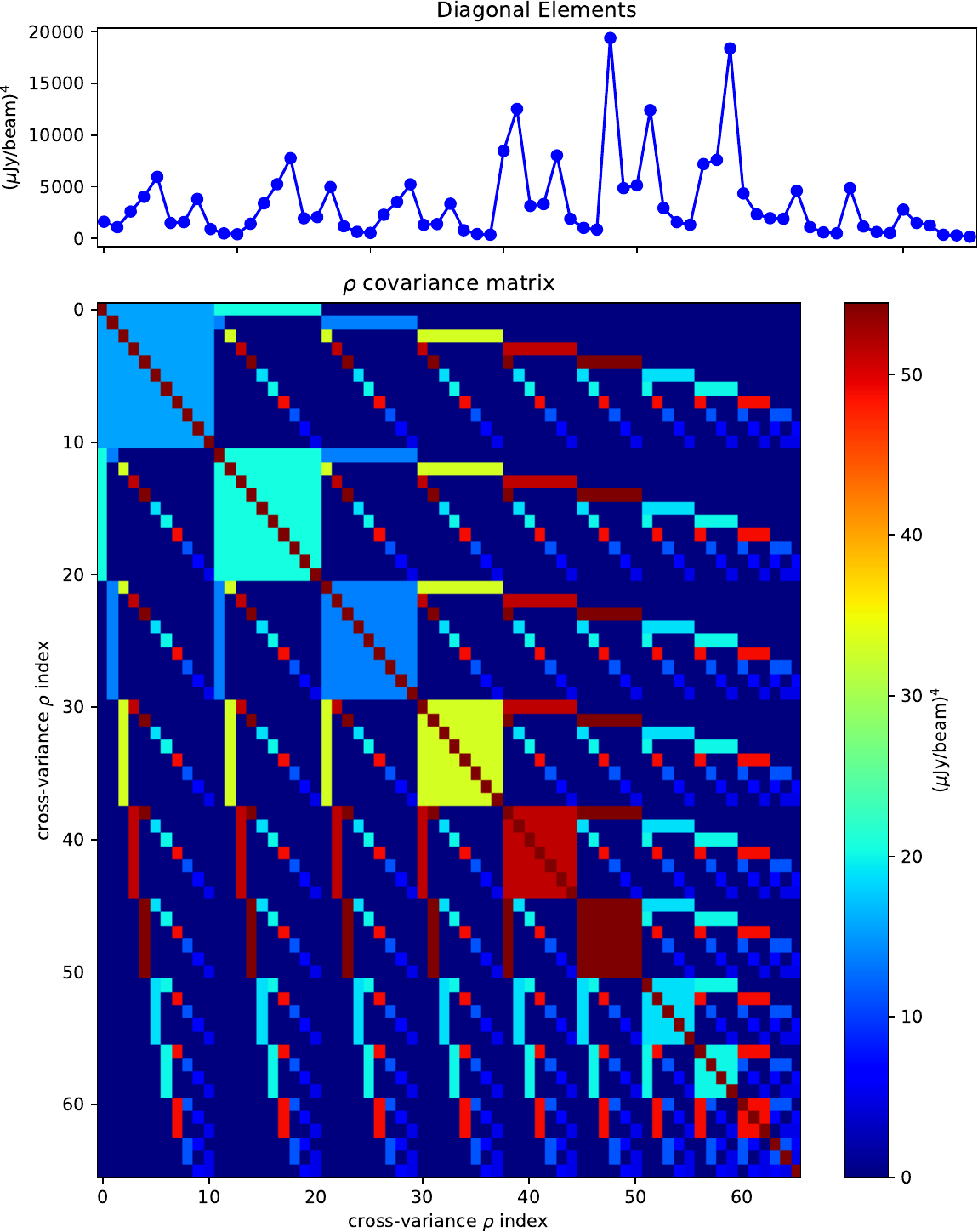}
\caption{Covariance matrix of the 66 \xrms\ estimates. Each pixel of this image is an element of this $66\times66$ matrix, and its value is given by the color bar. The color scale is saturated on purpose to display the off-diagonal terms. The top plot shows the diagonal elements of the matrix. The blocks, dots and lines that appear in the matrix reflect the correlations when the same map is used in Eq.~(\ref{eq:x_abcd_1}).}
\label{fig:rho_covar_matrix}
\end{figure}

In this work, the timeline processing proceeds per subscan, and they have different orientations on the sky (cf.~Sect.~\ref{se:gn}). As discussed in Sect.~\ref{se:transfer_function}, this leads to different values of $f$. Filtering also affects the noise. In general, filtering mostly affects low frequency components and therefore has less effect on the statistical noise than on the signal. The $1/f^2$ correction in the definition of $X_{AB,CD}$ is therefore somewhat conservative in terms of statistical uncertainty.\\
Figure~\ref{fig:all_rho} shows the 66 \xrms\ estimates at 1.2\,mm, and Fig.~\ref{fig:rho_covar_matrix} shows their associated covariance matrix.

Although we prefer to use likelihoods to derive our estimates of confusion and their associated confidence intervals, we here provide a direct way to estimate the overall \xrms\ for the sake of completeness. We have checked that in the high-S/N case of 1.2\,mm, the following derivation matches the maximum likelihood value given in the text.

We can gather all our \xrms\ estimates in a single vector $\{\rho_i\}_i$ and relabel the covariance matrix $X_{ij}$ terms accordingly. The optimal way to combine $\rho_i$ derives from the minimisation of the log likelihood

\begin{equation}
-2\ln \mathcal{L} = (\rho-\hat{\rho})^T X^{-1} (\rho-\hat{\rho})\,,
\end{equation}

thus,
\begin{eqnarray}
\frac{\partial (-2\ln \mathcal{L}) }{\partial \hat{\rho}} = 0 &\Leftrightarrow& \frac{\partial}{\partial \hat{\rho}} (\rho-\hat{\rho})^T X^{-1}(\rho-\hat{\rho}) = 0\nonumber\\
&\Leftrightarrow&
\frac{\partial}{\partial \hat{\rho}} \sum_{ij} \rho_i X^{-1}_{ij}\rho_j - \rho_i X^{-1}_{ij}\hat{\rho} - \hat{\rho}X^{-1}_{ij}\rho_j + \hat{\rho}^2X^{-1}_{ij} = 0 \nonumber\\
&\Leftrightarrow&
-2\sum_{ij}X^{-1}_{ij}\rho_j + 2\hat{\rho}\sum_{ij}X^{-1}_{ij} = 0 \nonumber \\
&\Leftrightarrow&
\hat{\rho} = \frac{\sum_{ij} X^{-1}_{ij}\rho_j}{\sum_{ij}X^{-1}_{ij}}\,, \label{eq:opt_rho}
\end{eqnarray}

and the final uncertainty of $\hat{\rho}$ reads

\begin{equation}
\sigma_{\hat{\rho}}^2 = \frac{1}{\sum_{ij}X^{-1}_{ij}}\,.
\label{eq:sigma_rho}
\end{equation}

\section{Note on photometry and conventions}
\label{se:appendix_convention}

The NIKA2 maps used in this work are calibrated in mJy/beam, and the absolute calibration is performed following \cite{perotto20}. In this convention, a point source of flux $\phi$ would show up on the map $m_p$ as a Gaussian with an amplitude equal to $\phi$, i.e. $m_p = \phi g_p$, with $g_p$ normalised such that their maximum is 1. In the limit of no noise and if the source falls exactly at the center of a map pixel, one could just read the value of this pixel to determine the flux. In practice, a weighted average of the pixel values $m_p$ is required, and a simple estimator of the flux of the source is:

\begin{equation}
    \hat{\phi} = \frac{1}{\sum_p g_p^2}\sum_p g_p m_p
\end{equation}

Under the assumption that the noise per pixel $\sigma_p$ is independent, the variance of this estimator is expressed as
\begin{equation}
    {\rm Var}(\hat{\phi}) = \left(\frac{1}{\sum_p g_p^2}\right)^2\sum_p g_p^2\sigma_p^2
.\end{equation}

Thus, under the assumption of uniform noise per pixel $\sigma$, the uncertainty on the flux of the source is simply:

\begin{equation}
    \sigma_\phi = \frac{\sigma}{\sqrt{\sum_p g_p^2}}
    \label{eq:rms_per_pix_to_rms_per_beam}
.\end{equation}

We can now relate the instrumental noise per beam to the overall variance of the unmasked pixels of our maps. Let us use the same notation as in the text to denote the value of a pixel $A_p$, its signal content $S_p$ and its noise $a_p$, then:

\begin{eqnarray}
\sigma^2 &=& \frac{1}{N}\sum_{p=1}^N(A_p-\bar{A})^2 \nonumber\\
&=&\frac{1}{N}\sum_{p=1}^N(S_p-\bar{S} + a_p-\bar{a})^2 \nonumber \\
&=&\frac{1}{N}\sum_{p=1}^N(S_p-\bar{S})^2 + \frac{1}{N}\sum_{p=1}^N(a_p-\bar{a})^2 \nonumber\\
&=&\sigma_c^2 + \sigma_a^2
.\end{eqnarray}

According to Eq.~(\ref{eq:rms_per_pix_to_rms_per_beam}), the instrumental noise per beam is now expressed as
\begin{equation}
    \sigma_{N/beam} = \sqrt{\frac{(\sigma^2-\sigma_c^2)}{\sum_p g_p^2}}
    \label{eq:instr_noise_per_beam}
\end{equation}

Computing $\sigma^2$ on our final maps and using the values of the confusion derived in Sect~\ref{se:results}, i.e. 139 and 39\,$\mu$Jy/beam, we thus find $\sigma_{N/beam} = 218$ and 60\,$\mu$Jy/beam at 1.2 and 2\,mm, respectively.

\section{The KID Möbius circle calibration method}
\label{app:cf}

In a nutshell, a KID is monitored by its response to two excitation signals, one in phase $I$ and the other one in quadrature $Q$. The KID tone, set at frequency $f$ is modulated by a known $\Delta f$ frequency shift, every half millisecond, that induces variations of these excitation signals by those measured quantities $dI$ and $dQ$~\citep{Catalano2014}. \citet{2013A&A...551L..12C} have shown different methods to use these quantities to monitor the variations of the resonance frequency $f_0$ of a KID when it collects photons, and how to relate them to the incoming power. We here present yet another method, that proves to be more robust against numerical integration. It relies on 1D (instead of 2D) polynomial fit. It is also easier to implement in algorithmic terms. The idea is to find how to project $I, Q$ and $dI$, $dQ$ onto an axis in a way that is as linear as possible with frequency, itself shown to be linear with optical power~\citep{Monfardini2014}. This is provided by the Möbius transform. 

We study the expected physical $Z=I+jQ$ complex dependency of a KID with frequency. \cite{2008ApPhL..93m4102G} have shown that 
\begin{equation}
Z=\frac{Z_{r}z_{1}}{Z_{r}z_{2}+z_{3}}\label{eq:circ1},
\end{equation}

where $z_{1}$, $z_{2}$ and $z_{3}$ are constant complexes, and
$Z_{r}=\frac{1}{2}+j(f-f_{0})/w$ where $f_{0}$ and $w$ are real values in Hz. The location of $Z$ is on a circle, at least approximately when near the resonance (see Fig.~\ref{fig:1dcircle2}.2 and the subsection below on Circle Fit). The inverse of $Z$ is a circle, but we can transform it to have an infinite radius circle {\it i.e.} a line, which can be expected to be linearised with the KID frequency. To simplify the solution of this inversion, we proceed in two steps. The first step is to scale, translate, rotate and reverse the initial circle so that it is identical to the reference circle defined as a $\frac{1}{2}$ radius, and $(\frac{1}{2},0)$ centre, in the complex plane, as defined by:
\[
Z_{ref}=\frac{1}{2}+\frac{1}{2}[\cos\phi+j\sin\phi]=\cos\frac{\phi}{2}\exp j\frac{\phi}{2}.
\]

The second step consists in inverting that reference circle (the Möbius transform), as $\frac{1}{Z_{ref}}=Z_{res}$ with
\[
Z_{res}=\frac{1}{\cos\frac{\phi}{2}}\exp-j\frac{\phi}{2}=1-j\tan\frac{\phi}{2}=x_{3}+j\,y_{3}\,,
\]

and $f=f_{0}+\frac{w}{2}\tan\frac{\phi}{2}$. The imaginary part of the inversion of the reference circle is just linearly dependent on the frequency, to first order. To calibrate this dependency (because the linearity is only approximate) we must rely on the $dI$, and $dQ$ measurements (see the subsection below on 1D polynomial fit).

\subsection*{Circle fit}

\begin{figure}
\centering
\includegraphics[angle=0,width=9cm]{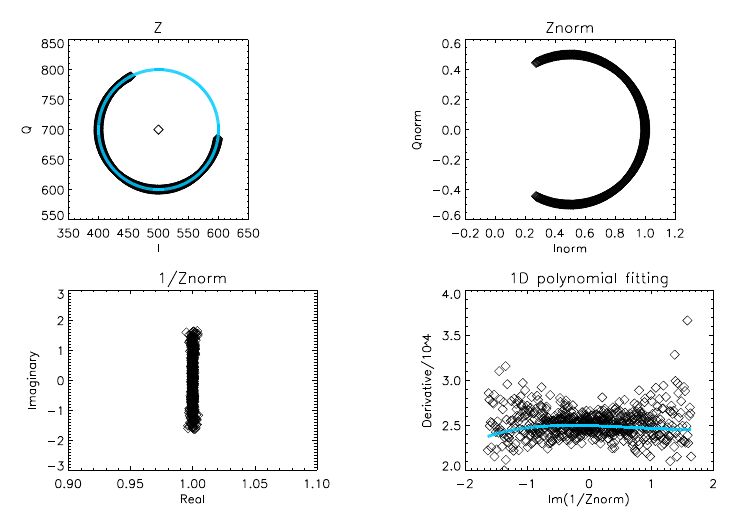}\label{fig:1dcircle1}
\caption{Illustration of our use of a Möbius transform to derive the total power measured by a KID on a simulation. Top-left: black diamonds are individual $I,\,Q$ points on a circle (blue curve). Top-right: normalised circle transformation of the points into the reference circle coordinates. The resonance is reached at (1,0). Bottom-left: Complex inverse of the points in the previous complex plane (the Möbius transform). Bottom-right: 1D polynomial fit (blue curve) to the derivative data points corresponding to $dI,\,dQ$ normalised, inverted and selected on the imaginary axis.}
\end{figure}

The first step is best achieved with a circle fit to the data samples\footnote{\url{R. Bullock, 2006, https://dtcenter.org/sites/default/files/community-code/met/docs/write-ups/circle_fit.pdf}}. Then, if $r$ is the radius of the circle and $x_{c},\,y_{c}$ its centre and $\alpha=\arctan(\frac{y_{c}}{x_{c}})$, the complex $Z$ becomes $Z_{norm}$ with $I_{norm}=-\frac{1}{2r}[(I-x_{c})\cos\alpha+(Q-y_{c})\sin\alpha]+\frac{1}{2}$,
and $Q_{norm}=\frac{1}{2r}[-(I-x_{c})\sin\alpha+(Q-y_{c})\cos\alpha]$.
This is shown in Fig.\,\ref{fig:1dcircle1}.1 where the top-left plot shows $Z$ and the top-right plot shows $Z_{norm}$. This figure is made with a simple simulation using Eq.\,(\ref{eq:circ1}) and including some noise. The bottom-left plot shows the inverse of $Z_{norm}$. The data points lie mostly along the imaginary line of abscissa unity and of ordinate $y_{3}$. The transform from mostly circular data on a plane to a one dimensional data stream is a particular case of Möbius transforms. They preserve angles, and this property is used in the following to propagate the calibration from the circle to the imaginary line.

\subsection*{1D polynomial fit}

\begin{figure}
\includegraphics[width=9cm]{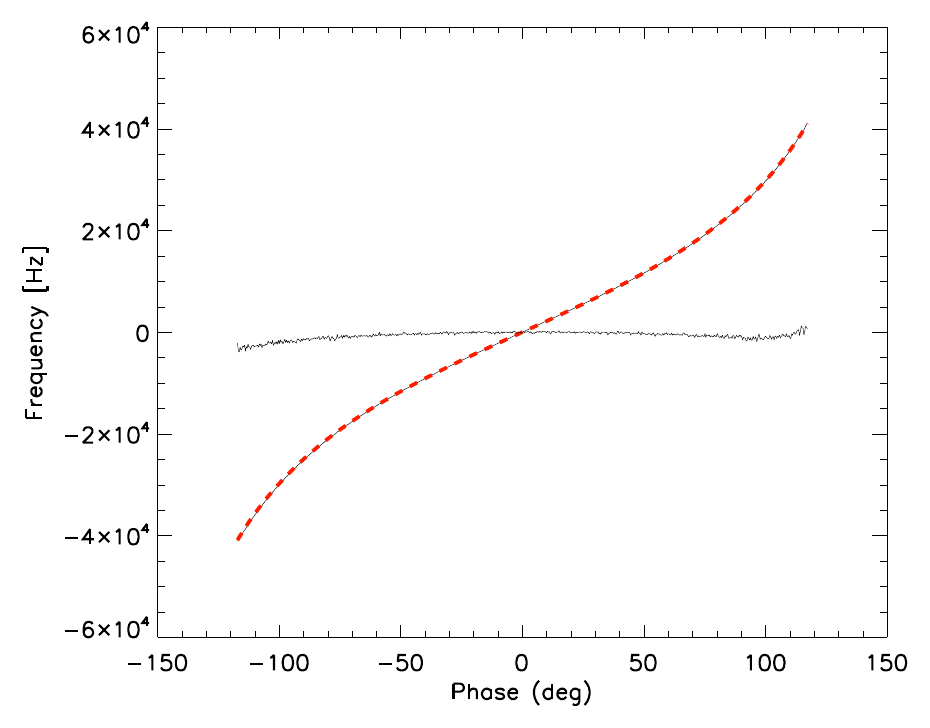}
\label{fig:1dcircle2}
\caption{Comparison between the KID simulation and the 1D polynomial retrieval. Relative KID frequency as a function of the phase $\phi$ in the circle, as deduced from the 1D polynomial fit (grey curve) and from the initial simulation (red dashed curve). The small wiggly horizontal black line represents the residual frequency (grey curve minus red curve) multiplied by a factor ten for visibility, as a function of the same phase.}
\end{figure}

We need to calibrate $y_{3}$ which is nearly proportional to
$f-f_{0}$. The modulation technique gives us this calibration via the derivative function. Let us call $dI$, and $dQ$, the measured variations of the signal with a modulation of frequency of $\Delta f$. The normalised variations are $dI_{norm}=-\frac{1}{2r}[dI\cos\alpha+dQ\sin\alpha]$, and $dQ_{norm}=\frac{1}{2r}[-dI\sin\alpha+dQ\cos\alpha]$. The derivative of the inverse complex is $dZ_{res}=-dZ_{norm}/Z_{norm}^{2}$. We then take the imaginary part of $dy_{3}$ and use it to calibrate $y_{3}$ in this way: we fit $\frac{\Delta f}{dy_{3}}=R_{n}(y_{3})$ where $R_{n}$ is a polynomial of degree $n$ (see bottom-right panel of Fig.\,\ref{fig:1dcircle1}.1) which is easy to integrate into $P_{n+1}$ (with $\dot{P}_{n+1}=R_{n}$). We then obtain the relative frequency of the KID from $y_{3}$ only with $f-f_{0}=P_{n+1}(y_{3})$
which is plotted in Fig.\,\ref{fig:1dcircle2}.2 along with the residual of the fit.

The method is robust if the polynomial fit is weighted. Indeed, the noise on the derivative measurement ($\frac{\Delta f}{dy_{3}}$) increases as the phase changes (see the bottom-right panel of Fig.\,\ref{fig:1dcircle1}.1 where the flaring away from the zero phase is apparent). The noise is proportional to the noise in $dy_{3}$ divided by $dy_{3}^{2}$. The noise in $dy_{3}$ is not constant. It can be shown that it is inversely proportional to the distance of the current $I, Q$ point to the point in the circle which is opposite to the resonance point.
This noise model can also be used to show that the noise on the recovered frequency is flaring as we go away from the resonance. To a good approximation, the rms noise is proportional to $1+\tan^{2}\frac{\phi}{2}=1+(\frac{2(f-f_{0})}{w})^2$. Hence, although a KID is a linear device, its noise depends on the distance of the measuring tone frequency to the resonance frequency.

This Möbius Circle calibration method is better at accounting for the quick variations of $dI, dQ$ and as such, is even more linear than previous methods. Another benefit of the method is that the sky noise, mostly due to water vapour fluctuations in the first kilometres of the atmosphere, is  calibrated more accurately between KIDs, on the long time scales, whereas other methods have a random drift $\sim \sqrt t$. Here, the global circle method ensures that the measured KID frequency is unique for a given point in the [$I$, $Q$] plane. The atmospheric common mode can thus be removed with less residual noise. 

\end{document}